\documentclass[english,aps,prb,showpacs,superscriptaddress,floats,amsmath,amssymb,floatfix,nobalancelastpage,twocolumn]{revtex4-2}
\usepackage[T1]{fontenc}
\usepackage[latin9]{inputenc}
\usepackage{amsmath}
\usepackage{amssymb}
\usepackage{graphicx}

\makeatletter

\newcommand*\LyXThinSpace{\,\hspace{0pt}}
\newcommand{\lyxdot}{.}

\usepackage{accents}

\makeatother

\usepackage{babel}
\begin{document}
\title{Finite temperature second harmonic generation in Kitaev magnets }
\author{Olesia Krupnitska}
\email{olesia.krupnitska@tu-braunschweig.de}

\affiliation{Institute for Theoretical Physics, Technical University Braunschweig,
D-38106 Braunschweig, Germany}
\affiliation{Institute for Condensed Matter Physics, National Academy of Sciences
of Ukraine, Svientsitskii Street 1, 790\LyXThinSpace 11, L'viv, Ukraine}
\author{Wolfram Brenig}
\email{w.brenig@tu-bs.de}

\affiliation{Institute for Theoretical Physics, Technical University Braunschweig,
D-38106 Braunschweig, Germany}
\begin{abstract}
We study electric field induced second harmonic generation (2HG) in
the Kitaev model. This frustrated magnet hosts a quantum spin-liquid,
featuring fractionalization in terms of mobile Majorana fermion and
static $\mathbb{Z}_{2}$ flux-vison elementary excitations. We show
that finite temperature 2HG allows to probe characteristic features
of both fractional quasiparticle types. In the homogeneous flux state
at low-temperatures, the 2HG susceptibility displays an oscillatory
spectrum, which is set by only the fermionic excitations and is subject
to temperature induced Fermi-blocking, generic to all higher harmonic
generation (HHG). In the intermediate to high temperature range, intrinsic
randomness, which emerges from thermally excited visons leads to drastic
changes of the 2HG susceptibility, resulting from resonance decoupling
over a wide range of energies. At the flux proliferation crossover,
we suggest an interpolation between these two temperature regimes.
Our results satisfy previously established symmetries for electric
field induced 2HG in Kitaev magnets. 
\end{abstract}
\maketitle

\section{Introduction}

Nonlinear optical (NLO) spectroscopy is a diverse tool of many interests.
In particular, second harmonic generation (2HG) has traditionally
served as a probe to study inversion symmetry breaking \citep{Boyd2008}.
Lately, this has played an eminent role in bulk graphene materials
\citep{Tokman2019,Vandelli2019,Ikeda2020,Zhang2022}, layered transition
metal dichalcogenides \citep{Janisch2014,Rosa2018,Ho2020,Kuang2021,Mouchliadis2021},
and transition metal monopnictide Weyl semimetals \citep{Wu2017}.
Probing inversion symmetry breaking by 2HG can even be related to
local properties \citep{Abulikemu2021}. Apart from 2HG, photovoltaic
shift currents, e.g., in ferroelectrics \citep{Baltz1981,Cook2017}
and topological insulators \citep{Tan2016} are another second order
NLO response of interest which is of similar sensitivity to symmetries.
Photogalvanic currents have been considered not only in the charge,
but also in the spin channel, e.g., for bilayer trihalides \citep{Ishizuka2022}.

Beyond exploring symmetries with low-order HG, the complete time-dependence
of the NLO response is of interest due to its relations to Floquet
group theory and Floquet topological insulators \citep{Alon1998,Morimoto2017,Neufeld2019}.
This is because the allowed emissions from higher harmonic generation
(HHG) can be analyzed in terms of certain spatiotemporal, dynamical
symmetries of time-periodically driven systems \citep{Alon1998,Ceccherini2001,Fregoso2013}

Lately, two-dimensional coherent NLO spectroscopy (2DCS), which is
a third-order echo response \citep{Kuehn2009,Kuehn2011}, has come
into focus, for accessing quasiparticle spectra and interactions.
This pertains not only to semiconductors \citep{Kuehn2011}, molecules
\citep{Engel2007}, and disordered many-body systems \citep{Mahmood2021}.
Instead, direct coupling of the driving external \emph{magnetic }fields
to the spin of correlated magnets has allowed to consider magnons
\citep{Lu2017}, kinks \citep{Wan2019}, spinons \citep{Li2021},
fractons \citep{Nandkishore2021}, Majorana fermions and visons \citep{Choi2020}
by 2DCS and also by a related second-order spectroscopy \citep{Qiang2023}.
Such applications have linked NLO spectroscopy to the very timely
topics of quantum spin-liquids (QSL) and fractionalization \citep{Savary2016}.

In the context of QSLs, the Kitaev magnet \citep{Kitaev2006} is of
particular interest \citep{Motome2019,Takagi2019}. This frustrated
magnet hosts a QSL due to an exact fractionalization of spin in terms
of two types of elementary excitations, namely mobile Majorana matter
and $\mathbb{Z}_{2}$ gauge flux which is localized in the absence
of external magnetic fields \citep{Kitaev2006}. It is realized by
an Ising model on the honeycomb lattice with bond-directional anisotropy
and may serve as a low-energy spin model for certain Mott-Hubbard
insulators with strong spin-orbit coupling \citep{Jackeli2009}. All
of its spin correlations are short ranged \citep{Baskaran2007} and
the flux-free sector allows for analytic treatment \citep{Kitaev2006}.
In a finite magnetic field the Kitaev magnet opens a gap, supporting
chiral edge modes \citep{Kitaev2006}. As for materials, $\alpha$-RuCl$_{3}$
\citep{Plumb2014} may currently be closest to representing the Kitaev
model, although additional exchange interactions lead to zigzag antiferromagentic
order below $7.1$K \citep{Cao2016}. This order can be suppressed
by in-plane magentic fields $H{\parallel}a$ \citep{Sears2017,Anja2017}
and leaves the most likely region for a low-temperature QSL in the
field range of $H\sim7{\dotsc}9$T \citep{Baek2017,Hentrich2018,Balz2019,Schonemann2020,Balz2021}.
Fractionalization in this QSL has been suggested to impact a multitude
of spectroscopic probes, including inelastic neutron scattering \citep{Knolle2014,Banerjee2016,Banerjee2017,Do2017},
Raman scattering \citep{Sandilands2015,Nasu2016,Wulferding2020},
resonant X-ray scattering \citep{Halasz2016}, phonon spectra \citep{Metavitsiadis2020,Metavitsiadis2022,Ye2020,Feng2021,Feng2022,Li_diff_2021},
and ultrasound propagation \citep{Hauspurg2023}.

Recently, in a first work \citep{Kanega2021}, the NLO response to
driving external \emph{electric} fields of a Kitaev magnet has been
added to the list of spectroscopies of fractional quasiparticles.
Based on a field-induced exchange-striction mechanism \citep{Katsura2009},
HHG by Majorana-fermions was shown to exist up to high order, using
the time-dependent density matrix from a Lindblad-equation aproach.
The latter approach was confined to a particular driving pulse-shape,
leaving spectral information about the higher harmonic (HH) susceptibilities
undisclosed. In addition, only the case of zero temperature was considered.
Therefore, since the magnetostrictive light-matter interaction does
not couple to $\mathbb{Z}_{2}$ flux directly, the impact of thermally
excited visons, i.e. the second type of fractional quasiparticles
of a Kitaev QSL on NLO spectroscopy remains to be understood.

Motivated by this, the purpose of this work is to study the effects
of finite temperature on the electric field driven NLO response of
a Kitaev QSL, focusing on the dynamical 2HG susceptibility. As prime
results, we show that not only Fermi-blocking occurs, but that visons
have a very strong impact, indicating that temperature is an additional
important parameter in NLO experiments. The paper is organized as
follows. In Sec. \ref{sec:model} we summarize the model. Sec. \ref{sec:N-th-harmonics}
details our evaluation of HHG and 2HG susceptibilities for homogeneous
and random gauge sectors, in Sec. \ref{subsec:LTNharm} and \ref{subsec:HT2harm},
respectively. Results and discussions are presented in Sec. \ref{sec:Results},
a summary in Sec. \ref{sec:Concl}. Additional information and further
calculations are deferred into App. \ref{app:A}-\ref{sec:Ut}.

\section{The Model \label{sec:model}}

We consider the Kitaev spin-model on the two dimensional honeycomb
lattice \citep{Kitaev2006}
\begin{equation}
H_{0}=\sum_{{\bf l},\alpha}J_{\alpha}S_{{\bf l}}^{\alpha}S_{{\bf l}+{\bf r}_{\alpha}}^{\alpha}\,,\label{eq:H0}
\end{equation}
where ${\bf l}=n_{1}{\bf R}_{1}+n_{2}{\bf R}_{2}$ runs over the sites
of the triangular lattice with ${\bf R}_{1[2]}=(1,0),\,[(\frac{1}{2},\frac{\sqrt{3}}{2})]$,
and ${\bf r}_{\alpha=x,y,z}=(\frac{1}{2},\frac{1}{2\sqrt{3}}),$ $(-\frac{1}{2},\frac{1}{2\sqrt{3}})$,
$(0,-\frac{1}{\sqrt{3}})$ refer to the basis sites $\alpha=x,y,z$,
tricoordinated to each lattice site of the honeycomb lattice with
Ising exchange $J_{\alpha}$, which we set isotropic in the absence
of electric fields, i.e., $J_{\alpha}=J$. While for $\alpha$-RuCl$_{3}$
most ab-initio studies suggest a sizable ferromagnetic Kitaev exchange
\citep{Takagi2019,Motome2019}, i.e. $J<0$ in Eq. (\ref{eq:H0}),
the sign of $J$ remains irrelevant in the absence of additional exchange
interactions or external magnetic fields. For the light-matter interaction
between the electric field $E$ and the spin system, we assume a minimal
dipole-coupling $-P\cdot E$, employing an exchange-striction mechanism
induced by orbital polarization \citep{Katsura2009,Lorenzana1995,Jurecka2000,Tokura2014}
\begin{equation}
P=\frac{\partial H_{0}}{\partial E}=g\sum_{{\bf l}}(S_{{\bf l}}^{x}S_{{\bf l}+{\bf r}_{x}}^{x}-S_{{\bf l}}^{y}S_{{\bf l}+{\bf r}_{y}}^{y})\,,\label{eq:P}
\end{equation}
where, to simplify symmetry matters, we set the field $\mathbf{E}=E\mathbf{e}_{\perp,z}$
to be perpendicular to the $z$-bonds \citep{Symmetry}. $P$ is the
effective \emph{polarization} operator and $g$ is the magnetoelectric
coupling constant. The size of $g$ remains an open question for $\alpha$-RuCl$_{3}$.
However, it has been argued, that for fields with $E\sim0.1-1$ MV/cm,
energies of $|g\,E|\sim0.01-0.1J$ can be reached \citep{Kanega2021}.

\begin{figure}
\centering{}\includegraphics[width=0.65\columnwidth]{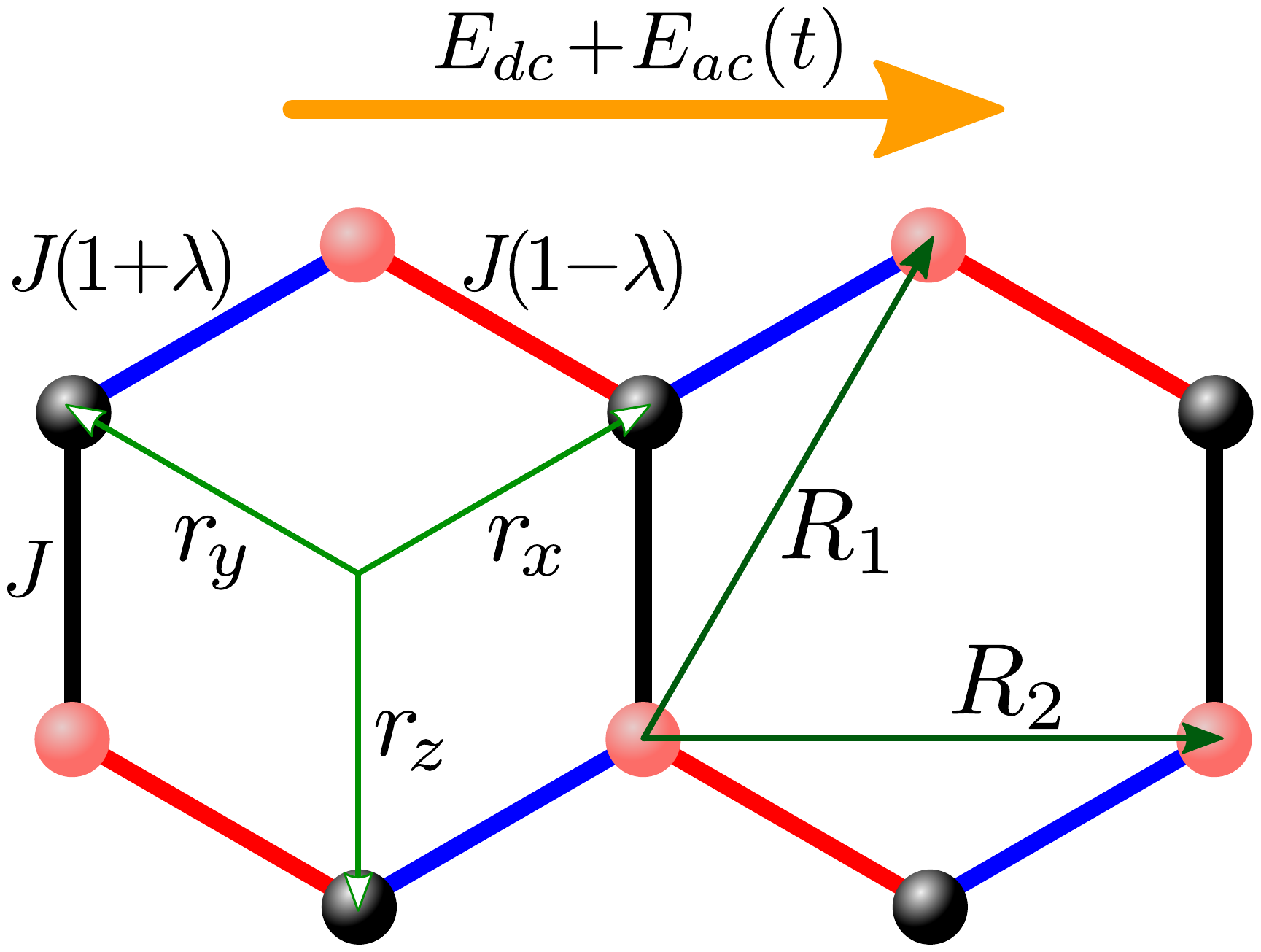}\caption{\label{fig:model} Kitaev model with (blue, red, black) $x$, $y$,
$z$-bonds, hosting $S_{{\bf l}}^{\alpha}S_{{\bf l}+{\bf r}_{\alpha}}^{\alpha}$exchange
with $\alpha{=}x$, $y$, $z$, respectively, in electric field $E_{dc}+E_{ac}(t)$
$\perp$ to $z$-bonds. $J\,[1+\lambda,1-\lambda,1${]} refers to
exchange interactions on $x$, $y$, $z$-bonds including dimerization
$\lambda=-gE_{dc}$ by static field.}
\end{figure}

The pure Kitaev model is invariant under the transformation $U$ of
reflection on the $z$-bond $(x{,}y){\rightarrow}({-}x{,}y)$, including
an exchange of spins $S^{x{,}y{,}z}{\rightarrow}(+{,}-{,}+)S^{y{,}x{,}z}$.
Both, polarization and electric field, change sign under $U$. Since
NLO susceptibilities at order $N$ of $E$ are rank-$(N{+}1)$ tensors
of $P$, see App. \ref{app:A}, this implies that even-$N$ response
vanishes unless the $U$-symmetry of $H$ is broken \citep{Kanega2021}.
To allow for such symmetry breaking, and for the remainder of this
work, we follow Ref. \citep{Kanega2021} and decompose $E=E_{dc}+E_{ac}(t)$
into a static (DC) and an dynamic (AC) part, the latter of which time-averages
to zero. As depicted in Fig. \ref{fig:model}, $E_{dc}$ can be absorbed
into a rescaled exchange $J_{\alpha}=J(1+\lambda,1-\lambda,1)$ with
$\lambda=-gE_{dc}$, thereby explicitly breaking the $U$-symmetry
of $H$. This procedure is reminiscent of the field-induced 2HG in
semiconductors \citep{Aktsipetrov1996} or graphene \citep{Bykov2012}.

Following established literature \citep{Kitaev2006,Motome2019}, Eqs.
(\ref{eq:H0},\ref{eq:P}) map onto a quadratic forms of Majorana
fermions in the presence of a static $\mathbb{Z}_{2}$ gauge $\eta_{{\bf l}}=\allowbreak\pm1$,
residing on, e.g., the $\alpha=\allowbreak z$ bonds
\begin{align}
\lefteqn{H_{0}-P(E_{dc}+E_{ac}(t))=H-PE_{ac}(t)s=}\label{eq:Ht}\\
 & -\frac{i}{2}\!\sum_{{\bf l},\alpha=x,y,z}\!\!\!\!J_{\alpha}\eta_{{\bf l},\alpha}\,a_{{\bf l}}c_{{\bf l}+{\bf r}_{\alpha}}+\frac{i}{2}\!\sum_{{\bf l},\alpha=x,y}\!\!\!\!\mathrm{sg}_{\alpha}\,a_{{\bf l}}c_{{\bf l}+{\bf r}_{\alpha}}\,gE_{ac}(t)\,,\nonumber 
\end{align}
where $\eta_{{\bf l},x(y)}=1$, $\eta_{{\bf l},z}=\eta_{{\bf l}}$,
and $\mathrm{sg}_{\alpha}=+(-)$ for $\alpha=x(y)$. There are two
types of Majorana particles, corresponding to the two basis sites.
We use the normalization $\{a_{{\bf l}},a_{{\bf l}'}\}=\delta_{{\bf l},{\bf l}'}$,
$\{c_{{\bf m}},c_{{\bf m}'}\}=\delta_{{\bf m},{\bf m}'}$, and $\{a_{{\bf l}},c_{{\bf m}}\}=0$.
Note that this Hamiltonian is diagonal in $\eta_{{\bf l},\alpha}$.
I.e., the optical exchange-striction mechanism does not excite $\mathbb{Z}_{2}$
fluxes. For the remainder of this work $\hbar=k_{B}=1$.

\section{HH susceptibilities\label{sec:N-th-harmonics}}

In this section we detail the evaluation of the HH response functions.
Two temperature regimes will be considered, namely $T\lesssim(\gtrsim)T^{\star}$,
where $T^{\star}\approx0.012\dotsc0.025J$ is the so-called flux proliferation
temperature. In a very narrow region of width $O(\pm0.01J)$ centered
around $T^{\star}$ the $\mathbb{Z}_{2}$-flux gets thermally populated,
changing the average link density rapidly from $\langle\eta_{\mathbf{l}}\rangle=1$
to $\langle\eta_{\mathbf{l}}\rangle=0$ \citep{Nasu2015,Metavitsiadis2017,Pidatella2019}.
In the following, we will employ this and divide the temperature axis
into approximately two regimes, i.e., the homogeneous flux sector
in Sec. \ref{subsec:LTNharm} for $T\lesssim T^{\star}$ and the random
flux sector \citep{RndSec} for $T\gtrsim T^{\star}$ in Sec. \ref{subsec:HT2harm}.
This approach has proven to work well on a quantitative level in several
studies of the thermal conductivity of Kitaev models \citep{Metavitsiadis2017,Pidatella2019,Metavitsiadis2017a}.

\subsection{Higher harmonic susceptibilties below $T^{\star}$\label{subsec:LTNharm}}

For $\eta_{{\bf l}}=1$ the Hamiltonian (\ref{eq:H0}) can be diagonalized
analytically in terms of complex Dirac fermions. This procedure has
been detailed extensively in various ways in the literature, e.g.,
\citep{Metavitsiadis2020,Metavitsiadis2021,Metavitsiadis2022,Motome2019}
and refs. therein. We merely state the quantities needed for the present
calculations. The real Majorana fermions are mapped onto complex ones
on \emph{half} of the momentum space by Fourier transforming $a_{{\bf k}}^{\phantom{\dagger}}=\sum_{{\bf l}}e^{-i{\bf k}\cdot{\bf l}}a_{{\bf l}}/\sqrt{N}$
with momentum ${\bf k}$ and similarly for $c_{{\bf k}}^{\phantom{\dagger}}$.
They satisfy $a_{{\bf k}}^{\dagger}=a_{-{\bf k}}^{\phantom{\dagger}}$.
Standard anticommutation relations apply, $\{a_{{\bf k}}^{\phantom{\dagger}},\allowbreak a_{{\bf k}'}^{\dagger}\}=\delta_{{\bf k},{\bf k}'}$,
$\{c_{{\bf k}}^{\phantom{\dagger}},\allowbreak c_{{\bf k}'}^{\dagger}\}=\delta_{{\bf k},{\bf k}'}$,
and $\{a_{{\bf k}}^{(\dagger)}c_{{\bf k}'}^{(\dagger)}\}=0$. The
diagonal form of $H$ is
\begin{equation}
H=\tilde{\sum_{{\bf k},\gamma=1,2}}\mathrm{sg}_{\gamma}\,\epsilon_{{\bf k}}\,d_{\gamma{\bf k}}^{\dagger}d_{\gamma{\bf k}}^{\phantom{\dagger}}\,,\label{eq:Hd}
\end{equation}
where $[c_{{\bf k}},a_{{\bf k}}]^{T}=\mathbf{u}(\mathbf{k})\,[d_{1{\bf k}},d_{2{\bf k}}]^{T}$
defines the quasiparticle fermions $d_{i{\bf k}}$ via a unitary transformation
$\mathbf{u}(\mathbf{k})$, App. \ref{sec:Ut}, and $\mathrm{sg}_{\gamma}$=1(-1)
for $\gamma$=1(2). The quasiparticles satisfy $\smash{d_{1(2){\bf k}}^{\dagger}=d_{2(1)-{\bf k}}^{\phantom{\dagger}}}$,
and $\smash{\tilde{\sum}}$ sums over half of momentum space. In cartesian
coordinates the quasiparticle energy $\epsilon_{{\bf k}}$ reads $\epsilon_{{\bf k}}=\allowbreak J[3+\allowbreak2\lambda^{2}+\allowbreak2(1-\lambda^{2})\cos(k_{x})+\allowbreak4\cos(k_{x}/2)\allowbreak\cos(\sqrt{3}k_{y}/2)-\allowbreak4\lambda\sin(k_{x}/2)\allowbreak\sin(\sqrt{3}k_{y}/2)]^{1/2}/2$.

Using the unitary trafo $\mathbf{u}(\mathbf{k})$, we may express
$P$ in terms of the quasiparticle Dirac fermions
\begin{equation}
P=g\tilde{\sum_{{\bf k},\mu\nu}}d_{\mu{\bf k}}^{\dagger}p_{\mu\nu}(\mathbf{k})d_{\nu{\bf k}}^{\phantom{\dagger}}\,,\label{eq:Pd}
\end{equation}
where, in cartesian coordinates, $p_{11}(\mathbf{k})=\allowbreak-p_{22}(\mathbf{k})=\allowbreak\sin(\allowbreak k_{x}/\allowbreak2)\allowbreak(2\lambda\sin(k_{x}/2)-\allowbreak\sin(\sqrt{3}k_{y}/2))/\allowbreak(2\epsilon_{{\bf k}})$
and $p_{12}(\mathbf{k})=\allowbreak p_{21}^{\star}(\mathbf{k})=\allowbreak-i\sin(k_{x}/2)\allowbreak(2\cos(\allowbreak k_{x}/\allowbreak2)+\allowbreak\cos(\allowbreak\sqrt{3}k_{y}/2))/\allowbreak(2\epsilon_{{\bf k}})$.
Obviously $P$ is not diagonal in the quasiparticle basis, implying
both, inter- and intraband excitations to occur.

\begin{figure}
\centering{}\includegraphics[width=0.9\columnwidth]{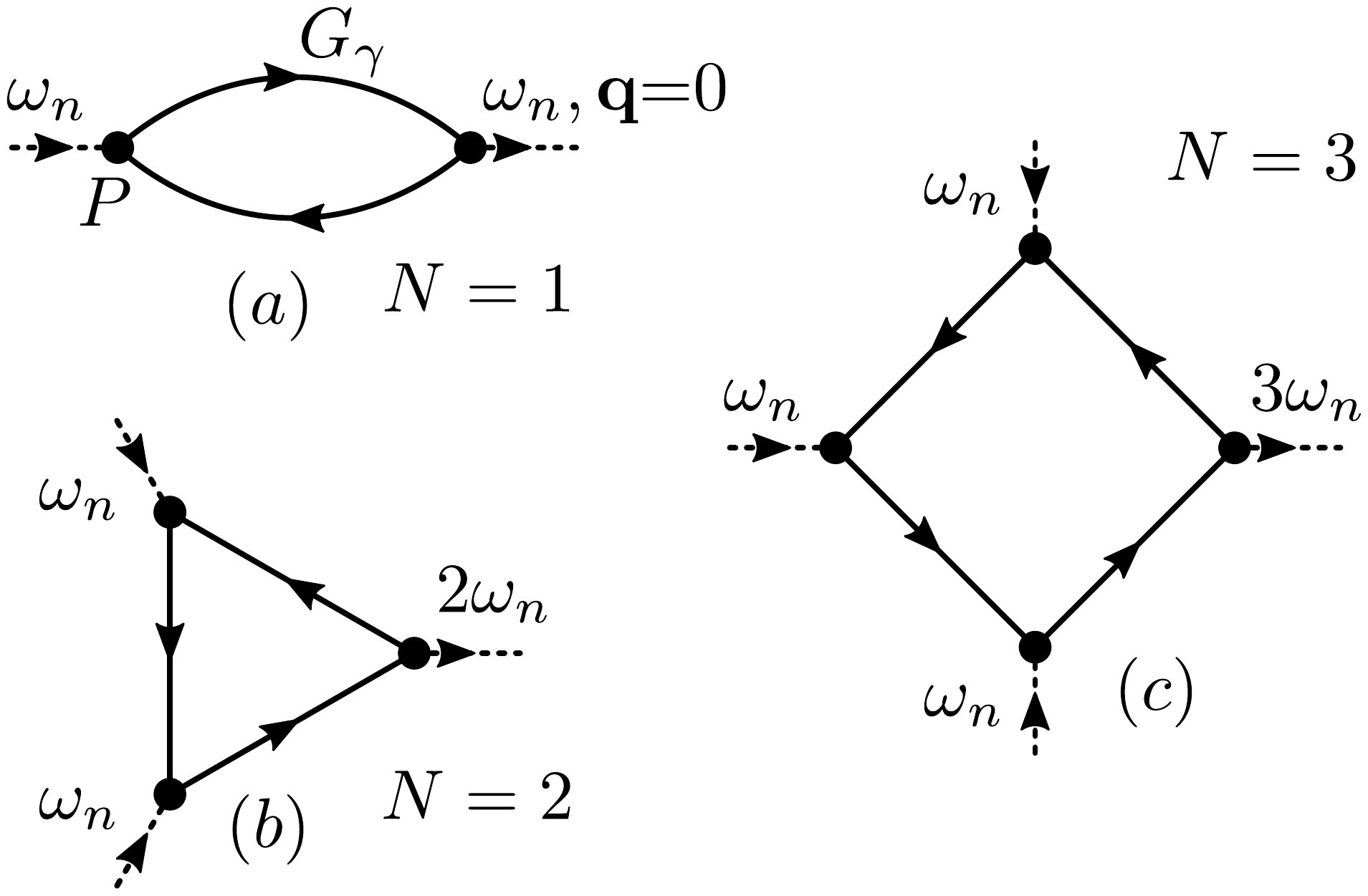}\caption{\label{fig:chidiag} Diagrams for $N$HG susceptibilities $\chi_{N\omega}(\omega)$
to leading order in $E_{ac}(t)$ for $N=1,2,3$.}
\end{figure}

Using Appendix \ref{app:A}, we are now in a position to formulate
the $N$th harmonic susceptibilities diagrammatically as in Fig. \ref{fig:chidiag}.
For the AC field we use $E_{ac}(t)=(e^{i\omega t}+\allowbreak e^{-i\omega t})A$,
with amplitude $A$. To appreciate these graphs, we first note, that
in principle $N$th harmonics can arise from any combination of contributions
by the field $\propto\exp[(\sum_{l}\pm1_{l})it\omega]$, such that
the sum of all signs of input frequencies satisfy $\sum_{l}\pm1_{l}=N$.
In turn $N$th harmonics can be generated at order $E_{ac}^{M}(t)$
with $M=N+2m$, $m\in\mathbb{N}$. For the remainder of this work
we will consider only the leading order, i.e., $M=N$ \citep{leadord}.
Second, we emphasize that for the latter situation, the intrinsic
permutation from Eq. (\ref{eq:Iperm}) is the identity, i.e., only
a single graph has to be considered for each $N$. Third, for all
purposes the wave vector of the incoming light can be set to zero
$\mathbf{q}=0$. Finally, for each Green's function line in Fig. \ref{fig:chidiag},
two contributions $G_{\gamma}(\mathbf{k},\epsilon_{n})$ with $\gamma=1,2$
from the two quasiparticle bands arise, fixing also the matrix elements
$p_{\mu\nu}(\mathbf{k})$ distributed along each graph.

Evaluation of the susceptibilities is straightforward. For Fig. \ref{fig:chidiag}(a)-(c)
we obtain
\begin{align}
\chi_{\omega}(\omega)= & g^{2}\tilde{\sum_{\mathbf{k}}}\frac{4(2f_{\mathbf{k}}{-}1)\epsilon_{\mathbf{k}}\,|p_{12}(\mathbf{k})|^{2}}{(\omega_{+}^{2}-4\epsilon_{\mathbf{k}}^{2})}\label{eq:chi1}\\
\chi_{2\omega}(\omega)= & g^{3}\tilde{\sum_{\mathbf{k}}}\frac{6(2f_{\mathbf{k}}{-}1)\epsilon_{\mathbf{k}}^{2}\,p_{11}(\mathbf{k})|p_{12}(\mathbf{k})|^{2}}{(\omega_{+}^{2}-\epsilon_{\mathbf{k}}^{2})(\omega_{+}^{2}-4\epsilon_{\mathbf{k}}^{2})}\label{eq:chi2}\\
\chi_{3\omega}(\omega)= & g^{4}\tilde{\sum_{\mathbf{k}}}\left[\frac{8(2f_{\mathbf{k}}{-}1)|p_{12}(\mathbf{k})|^{2}\epsilon_{\mathbf{k}}}{(9\omega_{+}^{2}-4\epsilon_{\mathbf{k}}^{2})}\right.\label{eq:chi3}\\
 & \left.\times\frac{(|p_{12}(\mathbf{k})|^{2}(\omega_{+}^{2}-\epsilon_{\mathbf{k}}^{2})+p_{11}^{2}(\mathbf{k})(\omega_{+}^{2}+4\epsilon_{\mathbf{k}}^{2}))}{(\omega_{+}^{2}-\epsilon_{\mathbf{k}}^{2})(\omega_{+}^{2}-4\epsilon_{\mathbf{k}}^{2})}\right]\,.\nonumber 
\end{align}
Calculating the diagrams for $\chi_{N\omega}(\omega)$ can be cast
into symbolic algebra code swiftly, providing analytical expressions
with minimal resources, easily up to $N>10$ \citep{higo}. Therefore
in practice, real-time HH response can be obtained for any input pulse
shape without repetitive solutions of Lindblad-equations by convolution
with \emph{analytic} expressions for $\chi_{N\omega}(\omega)$.

The susceptibilities (\ref{eq:chi1}-\ref{eq:chi3}) display the resonance
structure, typical of $N$HG. In particular, for each $N$, resonant
enhancement occurs at integer fractions of $2\epsilon_{\mathbf{k}}$,
down to $\omega=2\epsilon_{\mathbf{k}}/N$, indicative of the cooperative
transition of $N$ photons of frequency $\omega$ at a fermion gap
of $2\epsilon_{\mathbf{k}}$.

Not only the pure Kitaev model, but also its homogeneous flux sector
satisfies the $U$-symmetry discussed in Sec. \ref{sec:model}. Indeed,
following the diagrams of Fig. \ref{fig:chidiag} and Eqs. (\ref{eq:chi1}-\ref{eq:chi3}),
it is clear, that any even-$N$ harmonic susceptibility, including
those at sub-leading order in $E_{ac}(t)$, contain an odd {[}even{]}
power of $p_{11}(\mathbf{k})$ {[}$|p_{12}(\mathbf{k})|${]}. Moreover,
for vanishing static fields, i.e., $\lambda=0$, $\epsilon_{k_{x},k_{y}}=\allowbreak\epsilon_{-k_{x},k_{y}}$,
$p_{11}(k_{x},k_{y})=\allowbreak-p_{11}(-k_{x},k_{y})$, and $|p_{12}(k_{x},k_{y})|=\allowbreak|p_{12}(-k_{x},k_{y})|$,
as well as an identical relation for $k_{y}\rightarrow\allowbreak-k_{y}$.
Therefore, at $\lambda=0$, all even-$N$ harmonic susceptibilities
vanish. This agrees with Ref. \citep{Kanega2021}.

\subsection{Second harmonic susceptibilities above $T^{\star}$\label{subsec:HT2harm}}

In phases with random flux enclosed, i.e., for $T>T^{\star}$, Hamiltonian
(\ref{eq:Ht}) loses translational invariance and we resort to a formulation
amenable to numerical treatment in real space. This approach has been
detailed in the literature \citep{Metavitsiadis2020,Metavitsiadis2021,Metavitsiadis2022}.
We only list those points necessary to clarify the calculation of
susceptibilities.

The Majorana fermions on the $2N$ sites in Eq. (\ref{eq:Ht}) are
encoded into a spinor $A_{\text{\ensuremath{\sigma}}}^{\dagger}=(a_{1}\dots\allowbreak a_{{\bf l}}\dots\allowbreak a_{N},\allowbreak c_{1}\dots\allowbreak c_{{\bf l}+{\bf r}_{x}}\dots c_{N})$.
This is mapped onto a spinor of complex Dirac fermions $D_{\sigma}^{\dagger}=(d_{1}^{\dagger}\dots d_{N}^{\dagger},d_{1}^{\phantom{\dagger}}\dots d_{N}^{\phantom{\dagger}})$
using a unitary (Fourier) transform ${\bf F}$. I.e., ${\bf D}={\bf F}{\bf A}$
where bold faced symbols refer to vectors or matrices. ${\bf F}$
is constructed from two disjoint $N\times N$ blocks $F_{\sigma\rho}^{i=1,2}=e^{-i{\bf k}_{\sigma}\cdot{\bf R}_{\rho}^{i}}/\sqrt{N}$,
with $\sigma,\rho=1\dotsc N$ and ${\bf R}_{\rho}^{i}={\bf l}$ and
${\bf l}+{\bf r}_{x}$, for $a$- and $c$-Majorana lattice sites,
respectively. $\mathbf{F}$ merely serves as a device to introduce
complex fermions. ${\bf k}$ is chosen such, that for each ${\bf k}$,
there exists one $-{\bf k}$, with ${\bf k}\neq-{\bf k}$. Finally,
for convenience, ${\bf F}$ is rearranged such as to associate the
$d_{1}^{\dagger}\dots d_{N}^{\dagger}$ with the $2\,(N/2)=N$ 'positive'
${\bf k}$-vectors. In terms of these notations and for any set of
values $\{\eta_{\mathbf{l}}\}$
\begin{equation}
H={\bf D}^{\dagger}{\bf h}\,{\bf D}/2\,,\hphantom{aaa}P=g\,{\bf D}^{\dagger}{\bf p}\,{\bf D}/2\,.\hphantom{aaa}\label{eq:HPmat}
\end{equation}
The $2N\times2N$ matrices $\mathbf{h}$ and $\mathbf{p}$ are in
general non-diagonal and contain particle number non-conserving terms.
Finally, at this stage, and for a fixed distribution of $\{\eta_{\mathbf{l}}\}$
numerics is invoked to obtain a Bogoliubov transformation $\mathbf{U}$,
diagonalizing $\mathbf{h}$, i.e., $({\bf U}\mathbf{h}{\bf U}^{\dagger})_{\rho\sigma}=\delta_{\rho\sigma}\epsilon_{\rho}$,
with $\epsilon_{\rho}=(\epsilon_{1}\dots\epsilon_{N},-\epsilon_{1}\dots-\epsilon_{N})$
and quasiparticles ${\bf S}=\allowbreak{\bf U}{\bf D}$, for which
$H=\sum_{\rho=1}^{2N}\epsilon_{\rho}S_{\rho}^{\dagger}S_{\rho}^{\phantom{\dagger}}/2$.
We stay within a Nambu-notation with $\rho=1\dotsc2N$, i.e., keeping
the particle- and hole-range of $S_{\rho}^{(\dagger)}$, because the
quasiparticle form of the polarization $P=g\,{\bf S}^{\dagger}{\bf m}\,{\bf S}/2$
is \emph{not} simultaneously diagonal with $H$. It remains particle-number
non-conserving.

The quasiparticle Green's function $G_{\alpha\beta}(\tau)=-\langle T_{\tau}(S_{\alpha}^{\phantom{\dagger}}\allowbreak S_{\beta}^{\dagger})\rangle$
in Matsubara frequency space reads
\begin{equation}
G_{\alpha\beta}(\varepsilon_{n})=\delta_{\alpha\beta}G_{\alpha}(\varepsilon_{n})=\delta_{\alpha\beta}/(i\varepsilon_{n}-\epsilon_{\alpha})\,,\label{eq:G0}
\end{equation}
with $\varepsilon_{n}=(2n+1)\pi T$. To appreciate this textbook equation
in the present context, we introduce a notation, connecting the particle-
and hole-indices of $S_{\rho}$, namely $\bar{\rho}=\rho\mp N$ for
$\rho\gtrless N$. With that, \emph{anomalous} Green's functions simply
fulfill $-\langle T_{\tau}(S_{\alpha}^{\text{\ensuremath{\dagger}}}S_{\beta}^{\dagger})\rangle=-\langle T_{\tau}(S_{\bar{\alpha}}^{\phantom{\dagger}}S_{\beta}^{\dagger})\rangle=G_{\bar{\alpha}\beta}(\tau)$,
with $\epsilon_{\bar{\alpha}}=-\epsilon_{\alpha}$. This renders normal
\emph{and} anomalous contractions in the diagrams of Fig. \ref{fig:chidiag}
straightforward, using Eq. (\ref{eq:G0}) only and summing proper
index combinations for the polarization vertices $m_{\alpha\beta}$.
Diagram Fig. \ref{fig:chidiag} (b) yields
\begin{align}
\lefteqn{\chi_{2i\omega_{n}}(i\omega_{n})=-g^{3}\,T\!\!\!\sum_{\alpha\beta\gamma,\varepsilon_{m}}[t_{\alpha\gamma}t_{\gamma\beta}t_{\beta\alpha}}\label{eq:c3rndtau}\\
 & \hphantom{aaaaaaa}\times G_{\alpha}(\varepsilon_{m}{+}2\omega_{n})G_{\beta}(\varepsilon_{m}{+}\omega_{n})G_{\gamma}(\varepsilon_{m})]\,,\nonumber 
\end{align}
with $t_{\alpha\beta}=(m_{\alpha\beta}-m_{\bar{\beta}\bar{\alpha}})/2$,
where the first addend refers to the normal contraction order, and
the second to the anomalous. Frequency summation and analytic continuation
results in
\begin{align}
\chi_{2\omega}(\omega)=g^{3}\sum_{\alpha\beta\gamma,\varepsilon_{m}} & \frac{t_{\alpha\gamma}t_{\gamma\beta}t_{\beta\alpha}}{2\omega_{+}-\epsilon_{\alpha}+\epsilon_{\gamma}}\left(\frac{f_{\beta}-f_{\alpha}}{\omega_{+}-\epsilon_{\alpha}+\epsilon_{\beta}}+\right.\nonumber \\
 & \left.\hphantom{aa}\frac{f_{\beta}-f_{\gamma}}{\omega_{+}-\epsilon_{\beta}+\epsilon_{\gamma}}\right)\,,\label{eq:c3rndomg}
\end{align}
with the Fermi function $f_{\alpha}=1/(\exp(\epsilon_{\alpha}/T)+1)$.
Obviously Eqs. (\ref{eq:c3rndtau},\ref{eq:c3rndomg}) can readily
be generalized to any HH susceptibility. We refrain from this.

To complete our evaluation of $\chi_{N\omega}(\omega)$ for $T\gtrsim T^{\star}$,
a sufficiently large number of random distributions $\{\eta_{{\bf l}}\}$
is generated, for each of which $\mathbf{U}$, $\epsilon_{\alpha}$,
$m_{\alpha\beta}$, and Eqs. (\ref{eq:c3rndomg}) are calculated numerically,
with a final average over all $\chi_{N\omega}(\omega)$ obtained \citep{C3symm}.

\section{Results and discussion\label{sec:Results}}

For the purpose of discussion and to highlight some of the results
of this work, it seems of help to sketch the physics by elementary
considerations on a two-level system. Since the homogeneous sector
is translationally invariant, all HHG processes can be viewed as occurring
on a disjoint collection of pairs of states $\{1\mathbf{k},2\mathbf{k}\}$
with energies $\{\epsilon_{\mathbf{k}},-\epsilon_{\mathbf{k}}\}$,
enumerated by $\mathbf{k}$. For discussion, we reduce this to a single
two-level Hamiltonian $H=\epsilon|1\rangle\langle1|-\allowbreak\epsilon|2\rangle\langle2|$,
an accompanying polarization $P=n|1\rangle\langle1|+\allowbreak m|2\rangle\langle2|+\allowbreak g|1\rangle\langle2|+\allowbreak g^{\star}|2\rangle\langle1|$,
and a driving field $E(t)=\allowbreak(e^{i\omega t}+\allowbreak e^{-i\omega t})A$,
with a combined Hamiltonian of $H+P\,E(t)$. We are interested in
the expectation value $\langle P\rangle_{\rho}(t)$, with respect
to the time-dependent density matrix $\rho(t)$. In the interaction
picture $\dot{\rho}(t)=\allowbreak-i[Q(t),\rho(t)]$, with $Q(t)=P(t)E(t)=e^{iHt}\allowbreak P\allowbreak e^{-iHt}\allowbreak E(t)=(n|1\rangle\langle1|+\allowbreak m|2\rangle\langle2|+\allowbreak ge^{2i\epsilon t}|1\rangle\langle2|+\allowbreak g^{\star}e^{-2i\epsilon t}\allowbreak|2\rangle\langle1|)\allowbreak E(t)$.

Without loss of generality, from the various commutator contributions
to $\rho(t)$ for $2$HG, analogous to App. \ref{app:A}, we pick
a single time-ordering, with all $Q(t)$ left of $\rho_{0}$, the
equilibrium density matrix at $t=-\infty$, set to the zero-temperature
limit $\rho_{0}=|2\rangle\langle2|$. Using only the $e^{-i\omega t}A$
component of the driving field for the purpose of $2$HG, the contribution
reads
\begin{align}
\rho_{2\omega}(t)=-A^{2} & \int_{-\infty}^{t}dt_{1}\int_{-\infty}^{t_{1}}dt_{2}\,ne^{-i\omega_{+}t_{1}}\,|1\rangle\langle1|\,\times\label{eq:1Ttau}\\
 & \hphantom{aa}ge^{-i(\omega_{+}-2\epsilon)t_{2}}\,|1\rangle\langle2|\,|2\rangle\langle2|+\dotsc\nonumber \\
=A^{2} & \,n\,g\,\frac{|1\rangle\langle1|\,|1\rangle\langle2|\,|2\rangle\langle2|}{(2\omega_{+}-2\epsilon)(\omega_{+}-2\epsilon)}e^{-i(2\omega-2\epsilon)t}+\dotsc\,.\nonumber 
\end{align}
The selected time ordering shows off in the sequence of projectors
$|\mu\rangle\langle\nu|$ and ``$\dotsc$'' refers to all orderings
discarded. From this density matrix $\langle P\rangle_{\rho}(t)=\allowbreak A^{2}\allowbreak e^{-i2\omega_{+}t}\allowbreak n|g|^{2}/\allowbreak((2\omega_{+}-2\epsilon)(\omega_{+}-2\epsilon))$.

Obviously, $\langle P\rangle_{\rho}(t)$ displays frequency doubling.
Moreover, the structure of matrix elements is consistent with Eq.
(\ref{eq:chi2}), replacing $n|g|^{2}\leftrightarrow\allowbreak p_{11}(\mathbf{k})|p_{12}(\mathbf{k})|^{2}$.
Finally, the sequence of projectors in the last line of Eq. (\ref{eq:1Ttau})
allows to interpret the resonance denominators: The first photon invokes
an interband transition with resonance $(\omega-2\epsilon)^{-1}$.
The second photon invokes an intraband transition with resonance $(2\omega-2\epsilon)^{-1}$.
Connecting this with Eqs. (\ref{eq:chi1}-\ref{eq:chi3}), an $N$HG
response function shows $N$ resonances, separated by fixed, \emph{coupled}
integer fractions of the two-level energy $\pm2\epsilon_{\mathbf{k}}$,
for each $\mathbf{k}$.

This physics is drastically altered by gauge disorder. It renders
the Hamiltonian diagonal in a set of single fermion states, the quantum
numbers of which will be completely mixed by the polarization operator.
The consecutive absorption of $m{=}1{\dotsc}N$ photons of Eq. (\ref{eq:1Ttau})
remains intact, however, the corresponding resonance denominators
$(m\omega_{+}-\epsilon_{\alpha}+\epsilon_{\beta})^{-1}$ are independently
distributed over all energies, i.e., they are \emph{decoupled}. The
transitions may be intra- or interband, depending on the sign of $\epsilon_{\alpha}\epsilon_{\beta}$.
This is the content of Eq. (\ref{eq:c3rndomg}) in terms of the non-diagonal
matrix elements $t_{\alpha\beta}$ and the resonance denominators.

\begin{figure}[tb]
\centering{}\includegraphics[width=0.85\columnwidth]{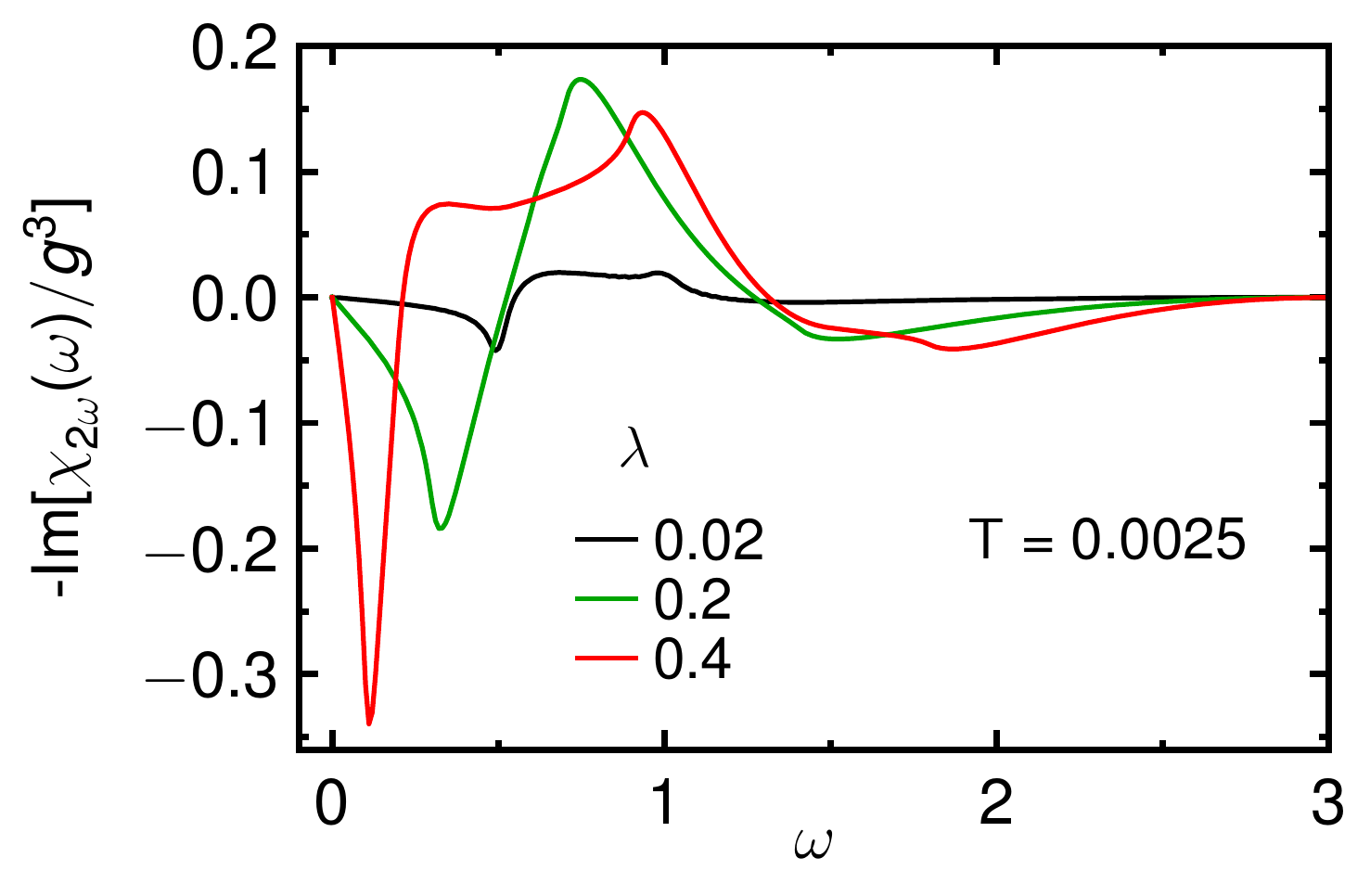}\caption{\label{fig:hmgvsedc}Imaginary part of dynamical $2$HG susceptibility
in homogeneous gauge sector versus $\omega$, for various static fields
$\lambda=-gE_{dc}$ at fixed temperature $T$. Linear system size
$L{=}400$, imaginary broadening $0.01$, energies in units of $J$. }
\end{figure}

In the homogeneous gauge sector, summing over $\mathbf{k}$, the product
structure of resonances of Eq. (\ref{eq:1Ttau}), comprising a sign
change between the poles, and interlocked by a fixed energy ratio
of $2$, can promote oscillations of $\mathrm{Im}\chi_{2\omega}(\omega)$,
in addition to oscillations which are induced by $p_{11}(\mathbf{k})$.
In random gauge sectors however, such oscillatory behavior should
be suppressed, in particular towards lower frequencies, where resonance
pairs from all energies overlap randomly. 

To substantiate the preceding, we now discuss several plots of $\chi_{2\omega}(\omega)$.
Fig. \ref{fig:hmgvsedc} displays the spectrum $\mathrm{Im}\chi_{2\omega}(\omega)$
from Eq. (\ref{eq:chi2}) at very low temperature $T=0.0025$ for
various $gE_{dc}$. Since $\chi_{2\omega}(\omega)$ is holomorphic
in the upper half plane, $\mathrm{Re}\chi_{2\omega}(\omega)$ follows
from Kramers-Kronig and will not be shown henceforth. The figure exemplifies
the 2HG selection rule, consistent with \citep{Kanega2021}. I.e.,
in the limit $gE_{dc}\rightarrow0$ the susceptibility vanishes. $\chi_{2\omega}(\omega)$
is antisymmetric with respect to $gE_{dc}$. The spectrum clearly
shows the oscillatory behavior discussed previously, exhibiting several
sign changes within the range $0<\omega<2\mathrm{max}(\epsilon_{\mathbf{k}})=3J$.
As is evident from the figure, the fermionic density-of-states modifications,
due to $gE_{dc}$ lead to characteristic shifts of van-Hove structures
in the spectrum. The values for $gE_{dc}$ in this figure are exploratory
only. Its experimentally accessible range remains to be clarified.

\begin{figure}[tb]
\centering{}\includegraphics[width=0.85\columnwidth]{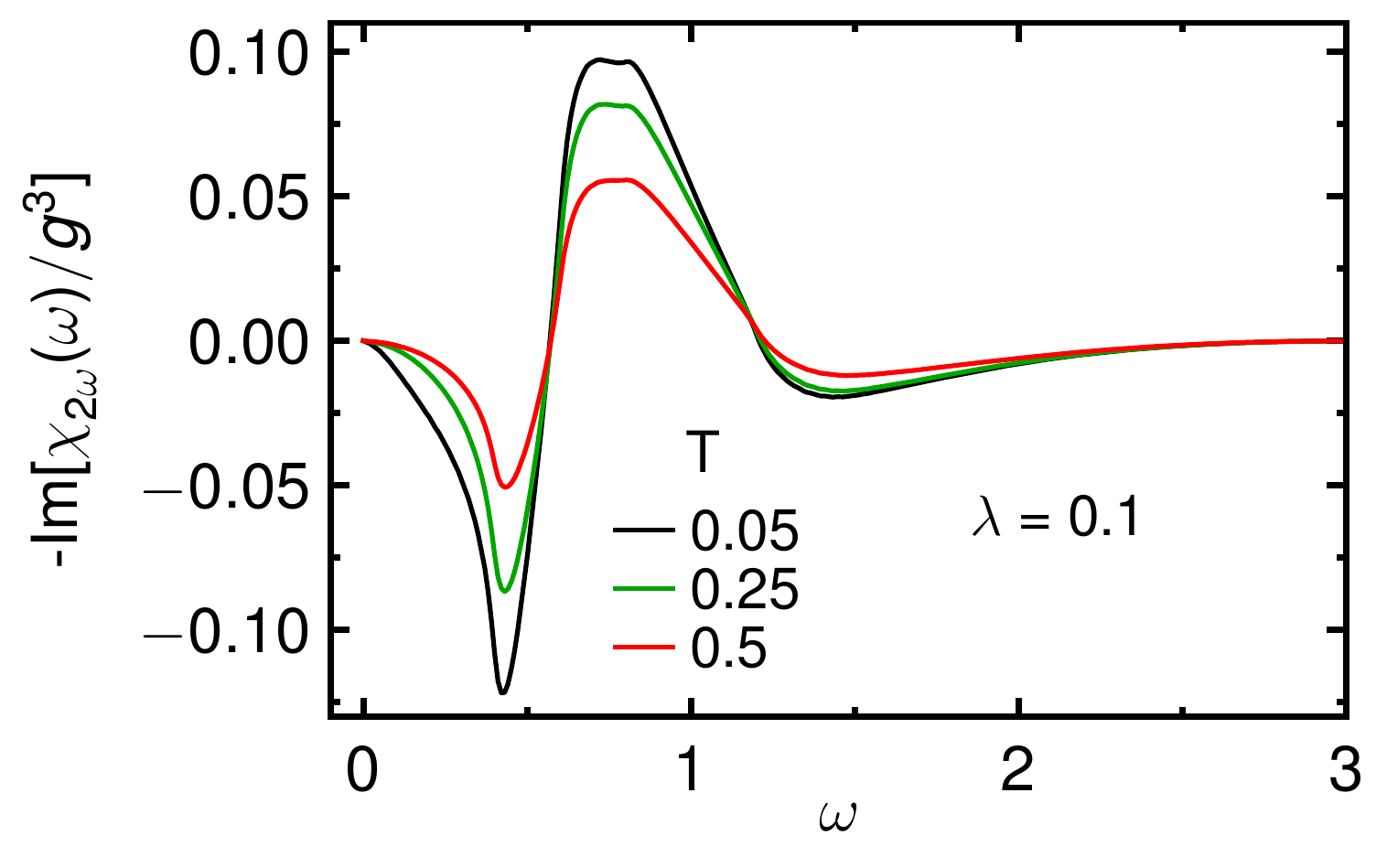}\caption{\label{fig:hmgvsT}Imaginary part of dynamical $2$HG susceptibility
in homogeneous gauge sector versus $\omega$, for various temperatures
$T$ at fixed static field $\lambda=-gE_{dc}$. Linear system size
$L{=}400$, imaginary broadening $0.01$, energies in units of $J$.}
\end{figure}

Next, in Fig. \ref{fig:hmgvsT}, we consider the temperature dependence
of the $2$HG susceptibility in the homogeneous gauge sector at a
finite $gE_{dc}$. Two of the temperatures displayed, i.e., $T=0.25J$
and $0.5J$ are well above the flux proliferation crossover. Analyzing
such temperatures with a homogeneous gauge is for demonstration only
and serves the purpose of clarifying the impact of thermally excited
flux later. The figure clearly shows the effect of the statistics
of the fermions. I.e., as the temperature increases, interband excitations
get blocked by thermal occupation, encoded by the factor $(1-2f_{\mathbf{k}})$
in Eqs. (\ref{eq:chi1}-\ref{eq:chi3}) for all HHG susceptibilities.
This so-called Fermi-blocking has been highlighted as a fingerprint
of fractionalization for a growing list of spectral probes of Kitaev
magnets, including Raman scattering \citep{Sandilands2015,Nasu2016,Wulferding2020},
resonant X-ray scattering \citep{Halasz2016}, phonon spectra \citep{Metavitsiadis2020,Metavitsiadis2022,Ye2020,Feng2021,Feng2022,Li_diff_2021},
and ultrasound propagation \citep{Hauspurg2023}. As a main result,
the present study adds HHG to this list.

\begin{figure}[tb]
\centering{}\includegraphics[width=0.85\columnwidth]{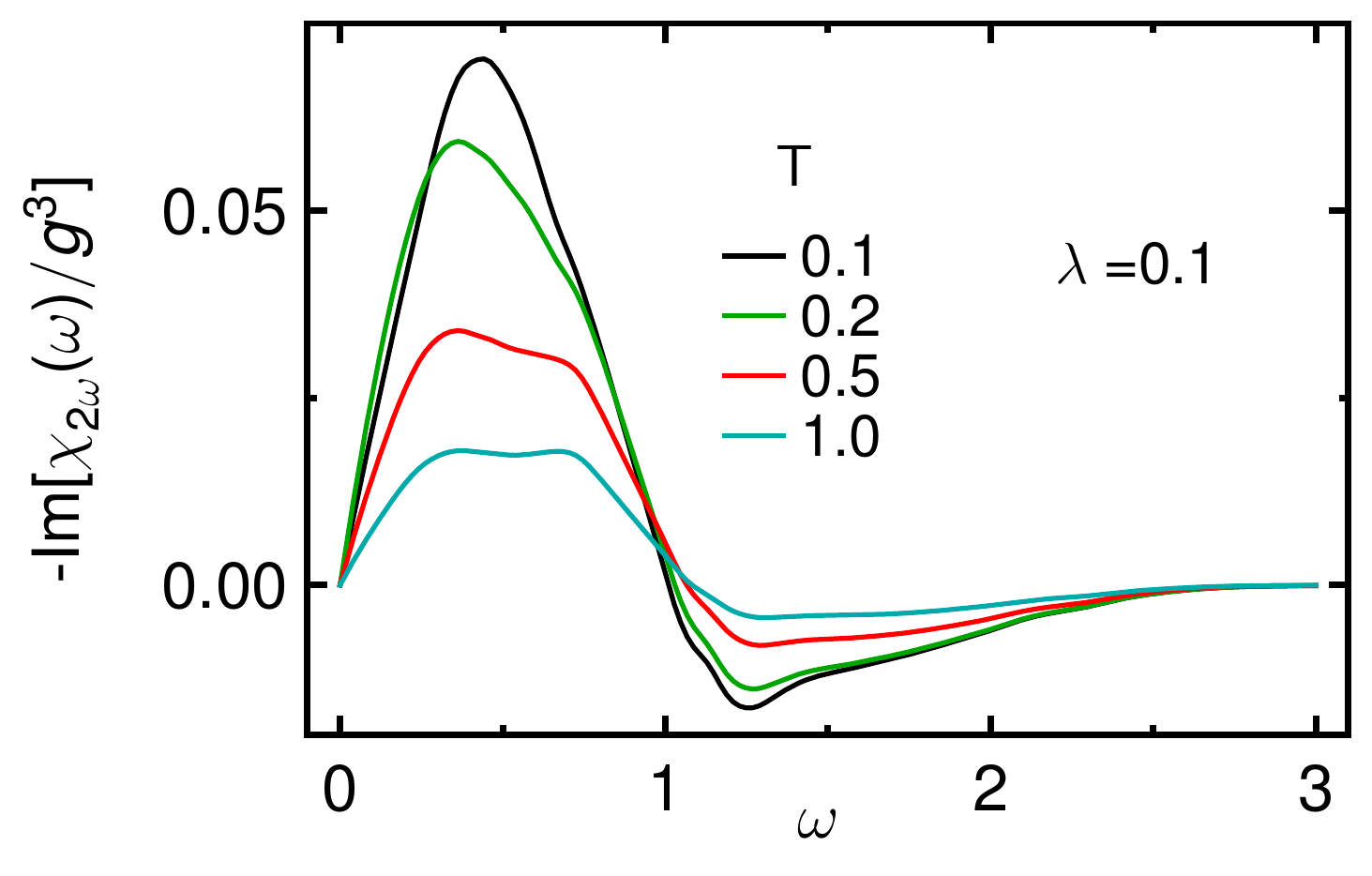}\caption{\label{fig:rnd05vsT}Imaginary part of dynamical $2$HG susceptibility
in random gauge sector versus $\omega$, for various temperatures
$T$ at fixed static field $\lambda=-gE_{dc}$. Linear system size
$L{=}30$, number of random realizations $62$, imaginary broadening
$0.05$, energies in units of $J$.}
\end{figure}

Fig. \ref{fig:rnd05vsT} displays the temperature dependence of the
$2$HG susceptibility in the random gauge sector, at the same fixed
static field, as in Fig. \ref{fig:hmgvsT}. First it should be noted,
that the linear system size $L=30$ is considerably smaller than for
the homogeneous sector. On the one hand, this is a prerequisite for
acceptable runtimes of Eq. (\ref{eq:c3rndomg}), comprising a Bogoliubov
transform, a matrix-product trace for each $\omega$, and an average
over $\{\eta_{\mathbf{l}}\}$ distributions. On the other hand, using
$L\sim30$ for a strictly homogeneous gauge is impractical, because
of finite-size degeneracies. These are absent in the random gauge
sector. Consequently, comparing $\chi_{2\omega}(\omega)$ between
homogeneous and random sectors, some quantitative finite-size differences
remain inevitable (see also App. \ref{app:RvsK}).

Apart from finite size effects, the spectra in Figs. \ref{fig:rnd05vsT}
and \ref{fig:hmgvsT} differ qualitatively. This difference can now
be understood in terms of the discussion at the start of this section.
I.e., in the random gauge sector the low- and high-frequency resonance
energies from Eq. (\ref{eq:c3rndomg}), $(\epsilon_{\alpha}-\epsilon_{\gamma})/2$
and $(\epsilon_{\alpha}-\epsilon_{\beta})$, $(\epsilon_{\beta}-\epsilon_{\gamma})$,
respectively, are distributed independent and randomly, with a DOS
that is almost flat \citep{Nasu2015} and they are coupled by non-diagonal
matrix elements. This modifies the lower energy region of $\chi_{2\omega}(\omega)$,
leading to significantly less oscillatory behavior as compared to
Fig. \ref{fig:hmgvsT}. In addition to this qualitative difference,
Fig. \ref{fig:rnd05vsT} shows Fermi-blocking similar to Fig. \ref{fig:hmgvsT}.
Therefore, as another main result of this study, for $T\gtrsim T^{\star}$,
both, the statistics of the fermionic quasiparticles, as well as the
visons impact the $2$HG susceptibility. 

\begin{figure}[tb]
\centering{}\includegraphics[width=0.85\columnwidth]{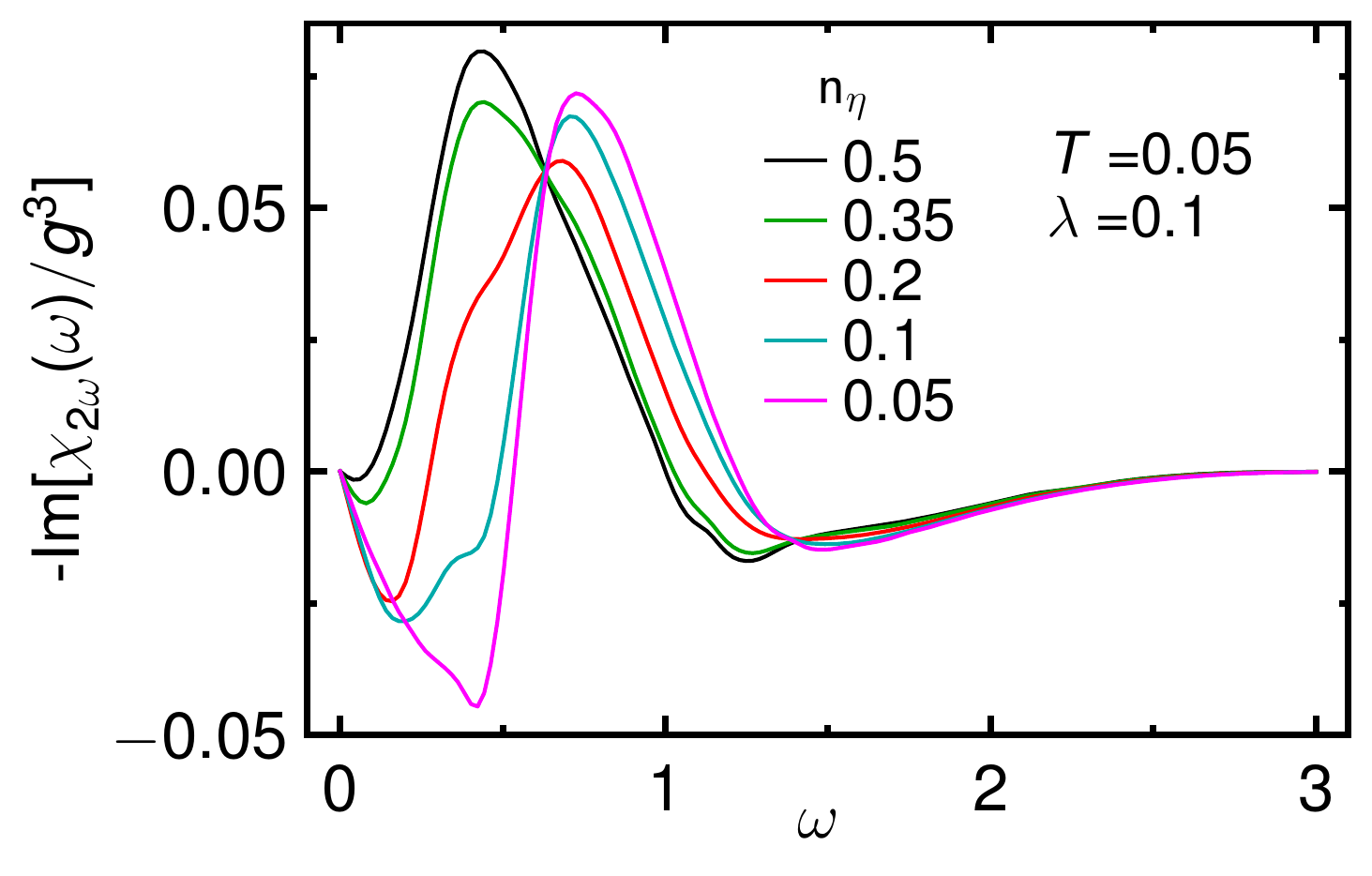}\caption{\label{fig:imvsfdens}Imaginary part of dynamical $2$HG susceptibility
versus $\omega$, for various flipped gauge link densities at fixed
temperature $T{=}0.05{\sim}T^{\star}$ and static field $\lambda=-gE_{dc}$.
Linear system size $L{=}30$, number of random realizations $62$,
imaginary broadening $0.05$, energies in units of $J$.}
\end{figure}

In Fig. \ref{fig:imvsfdens} we approximate the evolution of the $2$HG
susceptibility through the flux-pro\-li\-fe\-ra\-tion crossover.
An unbiased treatment of this requires evaluation of a dynamical 3-point
correlation-function at intermediate flux density, varying the temperature
across $T^{\star}$. Quantum Monte-Carlo \citep{Nasu2015} calculations
for this is an open issue. For exact diagonalization \citep{Metavitsiadis2017}
finite-size effects are expected to be detrimental. To make progress,
we therefore resort to a \emph{phenomenological} approach. This amounts
to fixing a temperature $T\approx T^{\star}$, e.g., $T=0.05J$ and
consecutively varying the average density $n_{\eta}$ of flipped gauge
links from a low value, essentially describing the homogeneous flux
sector, up to its maximum possible value at $n_{\eta}=1{/}2$. This
provides for an approximate interpolation. Since flux proliferation
occurs within a rather narrow temperature window of $O(T^{\star})\ll J$,
varying the temperature of the fermions can be discarded. First, comparing
the spectrum at $T=0.05$ in Fig. \ref{fig:hmgvsT} with that for
$n_{\eta}=0.05$ in Fig. \ref{fig:imvsfdens} provides a rough measure
for the finite size differences between $L=400$ and $30$. Otherwise,
these spectra show the same oscillatory behavior, representative of
the homogeneous sector. Using exactly $n_{\eta}=0$ within the $\mathbf{r}$-space
code is inconvenient because of large degeneracies at $L=30$. Remarkably,
and as another main result, Fig. \ref{fig:imvsfdens} corroborates
the discussion at the start of this section. I.e., the low-energy
modulations of $\chi_{2\omega}(\omega)$ are continuously removed
as $n_{\eta}$ increases. Quantitatively, the high energy spectrum
is less affected by the increase of flux-density.

\begin{figure}[tb]
\centering{}\includegraphics[width=0.85\columnwidth]{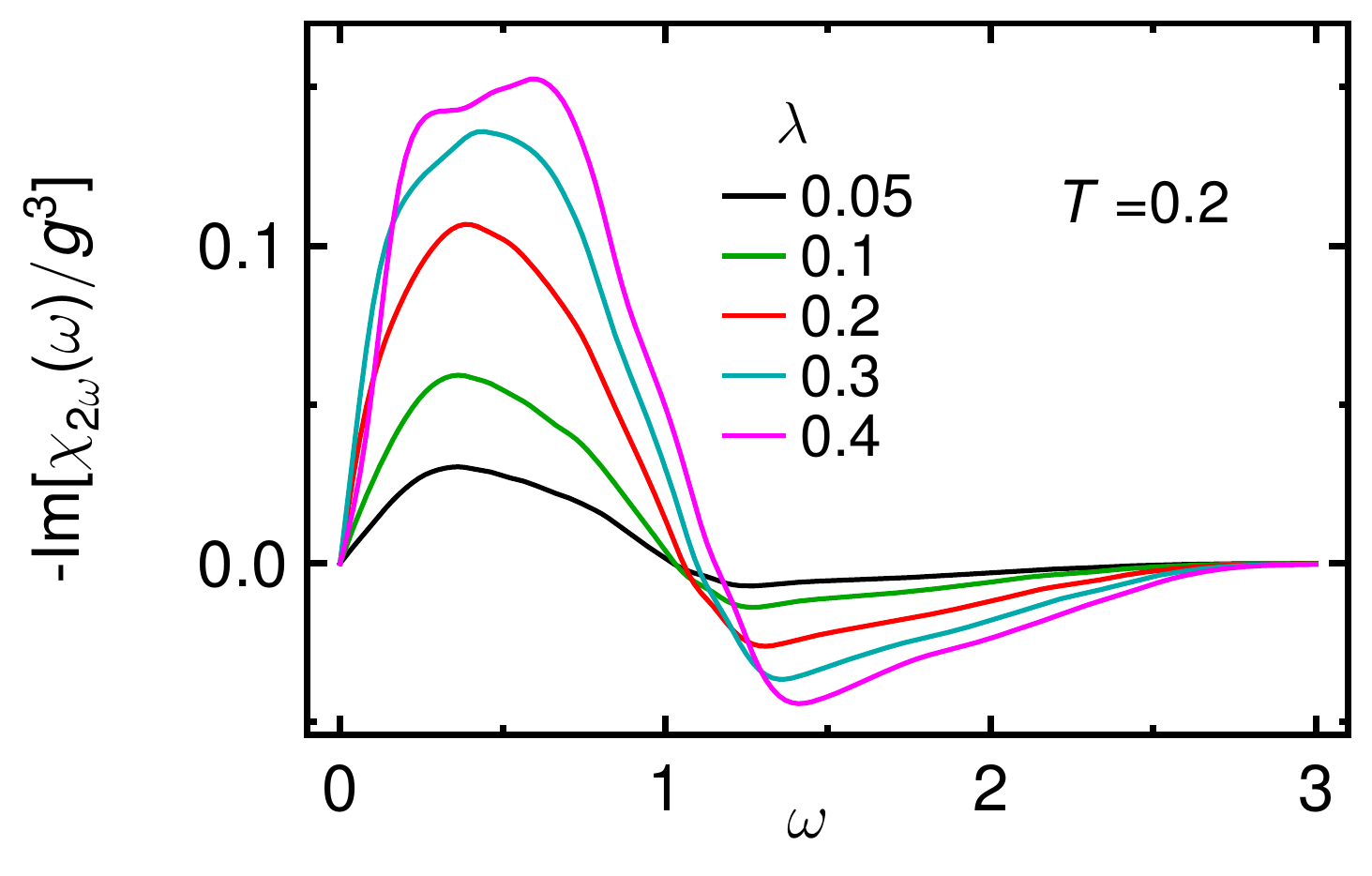}\caption{\label{fig:rndvsEdc}Imaginary part of dynamical $2$HG susceptibility
in random gauge sector versus $\omega$, for various static fields
$\lambda=-gE_{dc}$ at fixed temperature $T$. Linear system size
$L{=}30$, number of random realizations $62$, imaginary broadening
$0.05$, energies in units of $J$.}
\end{figure}

Finally, we show results for the dc-field dependence of $\chi_{2\omega}(\omega)$
in the random gauge sector in Fig. \ref{fig:rndvsEdc}. While for
a single gauge sector with a fixed random distribution of $\eta_{\mathbf{l}}$,
the $U$-symmetry of Sec. \ref{sec:model} will not be satisfied in
general, it is mandatory that after averaging over random $\eta_{\mathbf{l}}$
distributions, to within the statistical error $\chi_{2\omega}(\omega)$
must vanish for vanishing $gE_{dc}$, in order to encode the physics
of the pure Kitaev model. Fig. \ref{fig:rndvsEdc} clearly evidences
this behavior, implying that the pure Kitaev magnet shows no $2$HG
at any temperature.

\section{Summary\label{sec:Concl}}

We have studied finite temperature $2$HG in the Kitaev magnet. We
found that fractionalization in this frustrated quantum spin system
has a profound impact on the evolution of the 2HG susceptibility with
temperature. Mobile fermionic excitations, which are one kind of fractional
quasiparticles of this system, lead to an overall reduction of HHG
susceptibilities by Fermi-blocking on a temperature scale of the exchange
coupling constant. This is in line with other spectroscopic probes
of the Kitaev magnet. In addition however, a second low-temperature
scale $T^{\star}$ exists, in the narrow vicinity of which localized
$\mathbb{Z}_{2}$ visons, which are the second kind of fractional
quasiparticles, are thermally populated. This induces strong qualitative
changes of the 2HG susceptibility, by smoothing spectral oscillations
up to intermediate energies. In turn, both types of fractional quasiparticles
play an important role in finite temperature 2HG. While we have analyzed
the effects of the visons on the 2HG only, it is tempting to suggest
that this physics applies to all HHG.

Speculating on experimental consequences for the proximate Kitaev
magnet $\alpha$-RuCl$_{3}$, it should be realized first, that full
access to the conclusions of this work would be possible only in the
potential QSL phase in the in-plane field range of $H{\parallel}a\sim7{\dotsc}9$T.
Second, since $J\sim90K$ is in the terahertz range, the response
of convoluting few-cycle terahertz pulses with the 2HG susceptibility
should be very sensitive to the drastic changes between Figs. \ref{fig:hmgvsT}
and \ref{fig:rnd05vsT}. Therefore, apart from the gradual intensity
increase on a scale of $\sim90K$ by Fermi-blocking as $T$ is lowered,
a vison-induced strong intensity change should occur for $T\sim1\dotsc5K$.
Finally, the even HG selection rule seems an interesting case where
strain experiments could be of interest.

Our considerations have left aside the role of static magnetic fields
$H$ for the fermions and visons. For the former, magnetic field induced
gaps are low-energy features only and are smeared out by vison disorder
for $T>T^{\star}$. For the latter, vison dispersion generated by
magnetic fields could lead to interesting phenomena, which are beyond
the scope of this study.

Finally we note, that the methods described in this work are directly
applicable to the analysis of other types of higher order spectroscopies
in Kitaev magnets \citep{wbunp}. 
\begin{acknowledgments}
\emph{Acknowledgments}: Critical reading by Erik Wagner is acknowledged.
Work of W.B. has been supported in part by the DFG through Project
A02 of SFB 1143 (project-id 247310070). W.B. acknowledges kind hospitality
of the PSM, Dresden.
\end{acknowledgments}

\appendix

\section{$n$th order response diagrams\label{app:A}}

Let
\begin{equation}
H(t)=H-P\,E(t)\,,
\end{equation}
be a time dependent Hamiltonian. For labeling purpose, $P$ is dubbed
a ``polarization'' and $E(t)$ a time-dependent \emph{``}electric
field''. We assume the electric field to be adiabatically switched
on, starting at $t=-\infty$. The response at time $t$ of $P$ to
the perturbation $P\,E(t)$ is obtained from expanding the time-dependent
density matrix $\rho(t)$ in the interaction picture \citep{Fetter1971}\begin{widetext}
\begin{align}
\langle P\rangle(t)=&\langle P\rangle+ i \int_{-\infty}^{t}\langle[P(t),P(t_{1})]\rangle E(t_{1})\,dt_{1}
+ i^{2}\int_{-\infty}^{t}\int_{-\infty}^{t_{1}}
\langle[[P(t),P(t_{1})],P(t_{2})]\rangle
E(t_{1})E(t_{2})\,dt_{1}dt_{2}
\nonumber \\
&+ i^{n}\int_{-\infty}^{t}\dots\int_{-\infty}^{t_{n}}
\langle[\dots[P(t),P(t_{1})],\dots,P(t_{n})]\rangle E(t_{1})\dots E(t_{n})\,dt_{1}\dots dt_{n}+\dotsc\,,
\label{res1}
\end{align}
where $P(t)=e^{iHt}Pe^{-iHt}$ is the time dependence within the interaction picture. Eq.(\ref{res1}) can be written in terms of retarded susceptibilities
\begin{align}
\Delta\langle P\rangle(t)=&\int_{-\infty}^{\infty}\chi(t,t_{1})\,E(t_{1})\,dt_{1}+\int_{-\infty}^{\infty}\int_{-\infty}^{\infty}\chi(t,t_{1},t_{2})\,E(t_{1})E(t_{2})\,dt_{1}dt_{2}+\dots
\nonumber \\
&+\int_{-\infty}^{\infty}\dots\int_{-\infty}^{\infty}\chi(t,t_{1},\dots,t_{2})\,E(t_{1})\dots E(t_{n})\,dt_{1}\dots dt_{n}+\dotsc
\label{sus}
\end{align}
where
\begin{align}
\chi(t,t_{1})=& i\Theta(t-t_{1})\langle[P(t),P(t_{1})]\rangle\,,
\\
\chi(t,t_{1},t_{2})=& i^2 \Theta(t-t_{1})
\Theta(t_{1}-t_{2})
\langle[[P(t),P(t_{1})],P(t_{2})]\rangle\,, \label{c2def}
\\
\chi(t,t_{1},t_{2},\dotsc,t_{n})=& i^{n}\Theta(t-t_{1})\Theta(t_1-t_2)\dotsc\Theta(t_{n-1}-t_{n})
\langle[\dots[P(t),P(t_{1})],\dotsc,P(t_{n})]\rangle
\end{align}
\end{widetext}$\chi(t,t_{1})$ is the standard 2-point linear response susceptibility.
As usual, since $H$ is time-independent, $\langle[P(t),\allowbreak P(t_{1})]\rangle$
can be recast into $\langle[P(t-t_{1}),P]\rangle$, highlighting the
dependence of $\chi(t,t_{1})$ on $t-t_{1}$.

All of the $n$-fold time integrations in Eq. (\ref{sus}) are totally
symmetric with respect to any permutation of the $n$ time arguments.
Therefore, all contributions to $\Delta\langle P\rangle(t)$ can be
accounted for by replacing all susceptibilities by their fully symmetric
part, dubbed intrinsic permutation symmetry \citep{Butcher1990}
\begin{align}
\lefteqn{\chi(t,t_{1},\dots,t_{n})\rightarrow\chi_{S}(t,t_{1},\dots,t_{n})}\nonumber \\
 & \phantom{aaaa}=\frac{1}{n!}\sum_{\pi}\chi(t,t_{\pi(1)},\dots,t_{\pi(n)})\,,\label{eq:Iperm}
\end{align}
where $\pi$ labels all permutations.

It is textbook knowledge \citep{Fetter1971}, that the Fourier transform
$\chi(\omega)$ at real frequencies $\omega$ of the retarted 2-point
function $\chi(t)=i\Theta(t)\langle[P(t),P]\rangle$ can be obtained
from the Fourier transform of the imaginary time 2-point function
$\chi(\tau)=\langle T_{\tau}(P(\tau)P)\rangle$ at Matsubara frequencies
$i\omega_{n}=i\,2\pi T\,n$ by analytic continuation onto the real
axis with $i\omega_{n}\rightarrow\omega+i0^{+}$. Beyond this however,
the analytic continuation procedure applies to the evaluation of \emph{any}
fully symmetrized retarded $n$-point functions of an interacting
systems for \emph{all} $n$, including the case of arbitrary vertex
operators, $P,P(t_{1}),\dots,P(t_{n-1})\rightarrow P_{0},P_{1}(t_{1}),\dots,P_{n-1}(t_{n-1})$
with $P_{i}\neq P_{j}$ for $i\neq j$. This has been proved in Refs.
\citep{Evans,Rostami}. Therefore, if $P$ can be expressed in terms
of Fermions/Bosons, standard diagrammatic methods can be applied to
calculate $\chi(t,t_{1},\allowbreak\dots,\allowbreak t_{n})$ for
any $n$.

\begin{figure}
\centering{}\includegraphics[width=0.8\columnwidth]{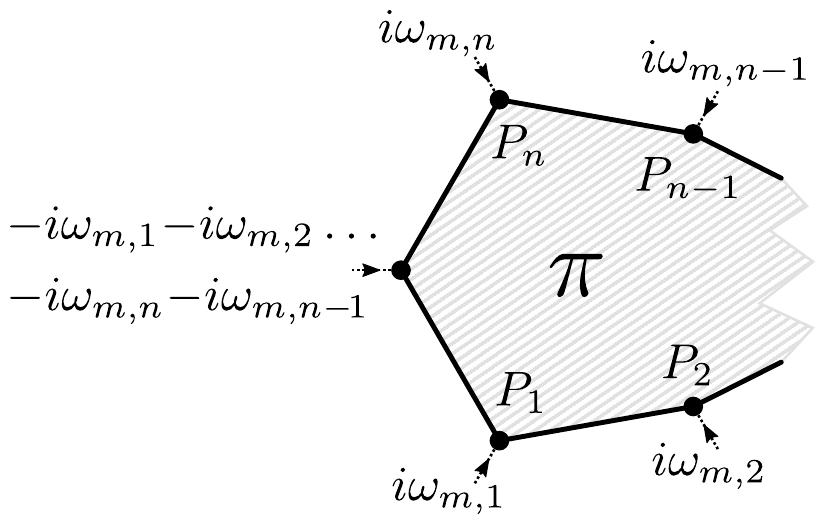}$\phantom{aaaaaa}$\caption{\label{fig:nptdiag}Intrinsically symmetrized $n+1$ point susceptibility.}
\end{figure}

Specifically, if $P$ is a quadratic form of Fermions/ Bosons, the
preceding implies the diagram Fig. \ref{fig:nptdiag} for the $n+\allowbreak1$
point susceptibility. Thick lines connecting the vertices $P_{j}$
refer to one-particle Green's functions and the hatched background
implies, that any kind of interactions may dress the graph, if $H$
provides for such. Finally, $\pi$ refers to an implicit sum over
all such graphs with the vertices $i\omega_{m,j}\allowbreak P_{j}$
permuted along the outer lines of the diagram. This clarifies the
correlation functions evaluated in Sec. \ref{subsec:LTNharm}.

\section{Equation of motion approach\label{app:EQM}}

For an approach alternative to the diagrams in Fig. \ref{fig:chidiag},
one can also evaluate the commutators in Eq. (\ref{res1}) directly
using the time dependent polarization operators within the interaction
picture. In the $\mathbf{k}$-space formulation for the homogeneous
state Eqs. (\ref{eq:Hd},\ref{eq:Pd}) yield 
\begin{equation}
P(t)=g\tilde{\sum_{{\bf k},\mu\nu}}e^{i(\epsilon_{\mu{\bf k}}-\epsilon_{\nu{\bf k}})t}d_{\mu{\bf k}}^{\dagger}p_{\mu\nu}(\mathbf{k})d_{\nu{\bf k}}^{\phantom{\dagger}}\,,\label{eq:pt}
\end{equation}
where $\epsilon_{\mu{\bf k}}=\mathrm{sg}_{\mu}\epsilon_{k}$, with
$\mathrm{sg}_{\mu}$ as defined after Eq. (\ref{eq:Hd}). Inserting
this into the commutators of Eq. (\ref{c2def}) we get
\begin{align}
\lefteqn{\langle[[P(t),P(t_{1})],P(t_{2})]\rangle=}\nonumber \\
 & \hphantom{aaa}g^{3}\tilde{\sum_{{\bf k}}}2p_{11}(\mathbf{k})|p_{12}(\mathbf{k})|^{2}(e^{i2\epsilon_{{\bf k}}(t_{1}-t_{2})}-e^{i2\epsilon_{{\bf k}}(t-t_{2})}+\nonumber \\
 & \hphantom{aaaaaaaaaa}e^{i2\epsilon_{{\bf k}}(t_{2}-t_{1})}-e^{i2\epsilon_{{\bf k}}(t_{2}-t)})(1-2f_{\mathbf{k}})\,,\label{eq:eqmK}
\end{align}
where the Fermi function $f_{\mathbf{k}}$, defined after Eq. (\ref{eq:chi3}),
results from thermal traces of type $\langle d_{\mu{\bf k}}^{\dagger}d_{\nu{\bf k}}^{\phantom{\dagger}}\rangle=\delta_{\mu\nu}\langle d_{\mu{\bf k}}^{\dagger}d_{\mu{\bf k}}^{\phantom{\dagger}}\rangle$.
Since $P(t)$ is quadratic in the fermions, all thermal traces in
Eq. (\ref{eq:eqmK}) are also. Inserting Eq. (\ref{eq:eqmK}) into
the $O(E^{2})$ addend of Eq. (\ref{res1}) using $E(t)=Ae^{-i\omega_{+}t}$
in order to obtain the 2HG response, we arrive at $\chi_{2\omega}(\omega)$
identical to Eq. (\ref{eq:chi2}). Completely analogous, following
the notation used in \ref{subsec:HT2harm}, the polarization expressed
in the interaction picture and in $\mathbf{r}$-space reads 
\begin{equation}
P(t)=\frac{g}{2}\sum_{\mu\nu}e^{i(\epsilon_{\mu}-\epsilon_{\nu})t}S_{\mu}^{+}m_{\mu\nu}S_{\nu}\,,\label{eq:rpt}
\end{equation}
where $\epsilon_{\mu}=(\epsilon_{1}\dots\epsilon_{N},-\epsilon_{1}\dots-\epsilon_{N})$
and $m_{\mu\nu}$ are matrix elements of $\mathbf{m}$, as defined
after Eq. (\ref{eq:HPmat}). Again, inserting this into the commutators
of Eq. (\ref{c2def}) we get 
\begin{align}
\lefteqn{\langle[[P(t),P(t_{1})],P(t_{2})]\rangle=g^{3}\sum_{\alpha\beta\gamma}t_{\alpha\gamma}t_{\gamma\beta}t_{\beta\alpha}\times}\nonumber \\
 & \hphantom{aaaa}\left(e^{i(\epsilon_{\gamma}-\epsilon_{\alpha})t}e^{i(\epsilon_{\beta}-\epsilon_{\gamma})t_{1}}e^{i(\epsilon_{\alpha}-\epsilon_{\beta})t_{2}}(f_{\beta}-f_{\alpha})+\right.\nonumber \\
 & \hphantom{aaaa}\left.e^{i(\epsilon_{\gamma}-\epsilon_{\alpha})t}e^{i(\epsilon_{\alpha}-\epsilon_{\beta})t_{1}}e^{i(\epsilon_{\beta}-\epsilon_{\gamma})t_{2}}(f_{\beta}-f_{\gamma})\right)\,,\label{eq:eqmR}
\end{align}
where $t_{\alpha\beta}$ are defined after Eq. (\ref{eq:c3rndtau}).
Performing the integrations in Eq. (\ref{res1}) similar to the $\mathbf{k}$-space
case, we arrive at $\chi_{2\omega}(\omega)$ identical to Eq. (\ref{eq:c3rndomg}).

\section{Real space versus momentum space calculations\label{app:RvsK}}

\begin{figure}[tb]
\centering{}\includegraphics[width=0.9\columnwidth]{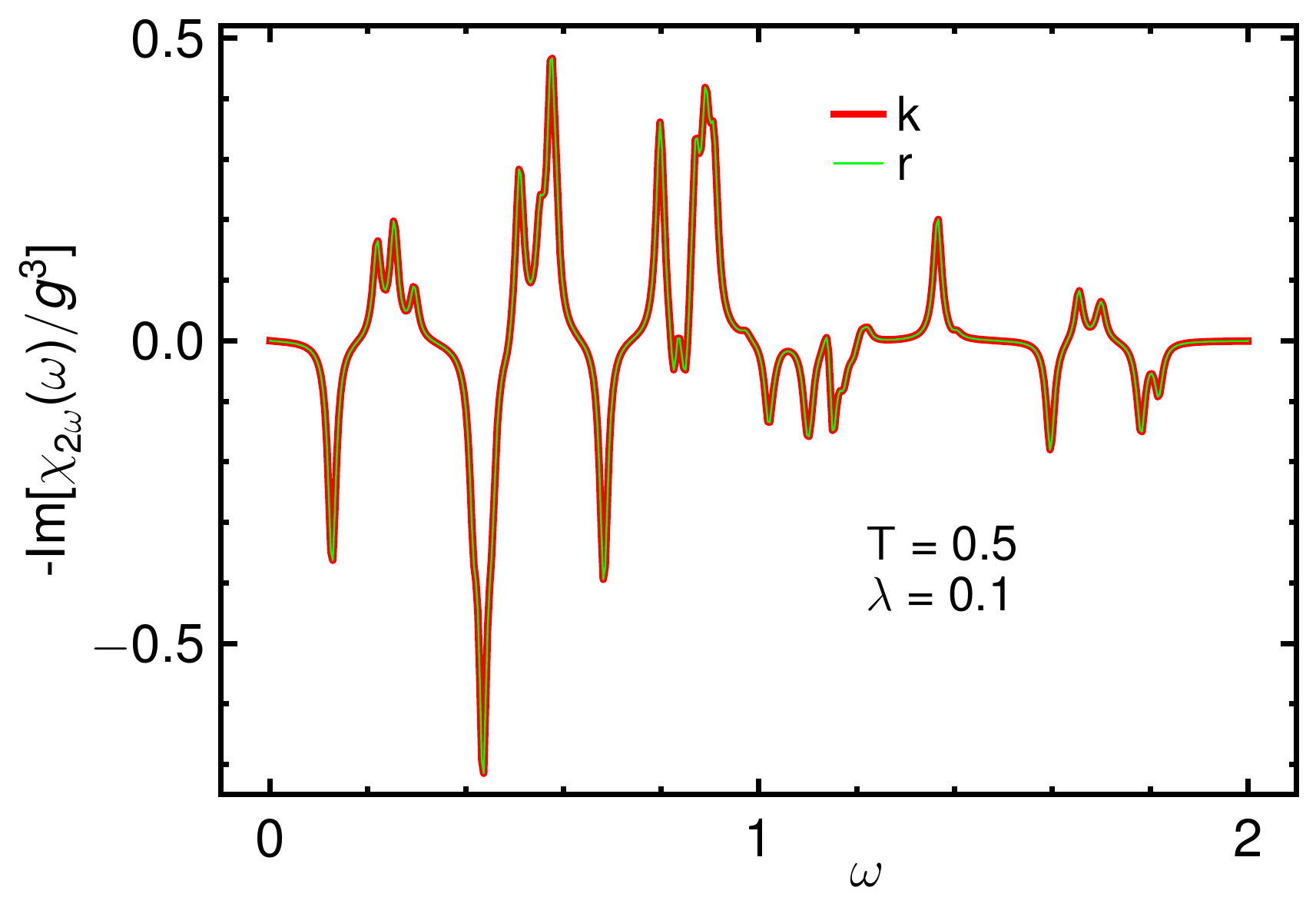}\caption{\label{fig:cmpHvsR}Section of imaginary part of dynamical $2$HG
susceptibility on very small system of linear size $L{=}8$, comparing
Eq. (\ref{eq:chi2}) (bold red) with fully numerical results from
Eq. (\ref{eq:c3rndomg}), forcing a homogeneous gauge (thin green)
and setting imaginary broadening $0.01$ to resolve single quasiparticle
poles. Energies in units of $J$.}
\end{figure}

This section is merely meant to prove numerically, that the two rather
diametric approaches used in Sec. \ref{subsec:LTNharm} and Sec. \ref{subsec:HT2harm},
i.e., the $\mathbf{k}$-space and $\mathbf{r}$-space calculations,
indeed yield identical results if used within the translationally
invariant, homogeneous gauge sector. For that purpose, we consider
a system, deliberately chosen small enough to resolve single quasiparticle
energies for a sufficiently small imaginary broadening and set the
temperature such as to involve both, positive and negative quasiparticle
energies. For such a case, we contrast $\mathrm{Im}\chi_{2\omega}(\omega)$
resulting from the analytical expressions Eq. (\ref{eq:chi2}) with
that obtained from the fully numerical procedure in $\mathbf{r}$-space.
A typical example is shown in Fig. \ref{fig:cmpHvsR}. The results
are identical to within numerical precision.

\section{Diagonalization of homogeneous sector\label{sec:Ut}\protect \\
}

Here, for completeness sake, we list the unitary transformation to
quasiparticles $\mathbf{u}(\mathbf{k})$, cited after Eq. (\ref{eq:Hd})
and known from the literature, e.g. \citep{Metavitsiadis2020}. The
quasiparticles are given by 
\begin{align}
\left[\begin{array}{c}
c_{{\bf k}}\\
a_{{\bf k}}
\end{array}\right] & =\left[\begin{array}{cc}
u_{11}({\bf k}) & u_{12}({\bf k})\\
u_{21}({\bf k}) & u_{22}({\bf k})
\end{array}\right]\left[\begin{array}{c}
d_{1{\bf k}}\\
d_{2{\bf k}}
\end{array}\right]\label{eq:4}\\
u_{11}({\bf k}) & =-u_{12}({\bf k})=\frac{i\sum_{\alpha}e^{-i{\bf k}\cdot{\bf r}_{\alpha}}}{2^{3/2}\epsilon_{{\bf k}}}\nonumber \\
u_{21}({\bf k}) & =u_{22}({\bf k})=\frac{1}{\sqrt{2}}\,,\nonumber 
\end{align}
where the momontum $\mathbf{k}$ is set by ${\bf k}=x\,{\bf G}_{1}+y\,{\bf G}_{2}$
with $x,y\in[0,2\pi[$, in terms of the basis ${\bf G}_{1[2]}=(1,-\frac{1}{\sqrt{3}})\,[(0,\frac{2}{\sqrt{3}})]$,
which is reciprocal to the triangular lattice basis listed after Eq.
(\ref{eq:H0}).

In terms of the reciprocal coordinates $x,y$, the quasiparticle energy
$\epsilon_{{\bf k}}$ stated after Eq. (\ref{eq:Hd}) reads $\epsilon_{{\bf k}}=J[3+2\lambda^{2}+\allowbreak2(1-\lambda^{2})\cos(x)+\allowbreak2(1-\lambda)\cos(x-y)+\allowbreak2(1+\lambda)\cos(y)]^{1/2}/2$.

Furthermore, the matrix elements of the dipole operator, cited after
Eq. (\ref{eq:Pd}), in reciprocal coordinates are $p_{11}(\mathbf{k})=-p_{22}(\mathbf{k})=(\cos(y)-\allowbreak\cos(x-y)+\allowbreak2\lambda(1-\allowbreak\cos(x)))/\allowbreak(4\epsilon_{{\bf k}})$
and $p_{12}(\mathbf{k})=p_{21}^{\star}(\mathbf{k})=-i(\sin(x-\allowbreak y)+\allowbreak2\sin(x)+\allowbreak\sin(y))/\allowbreak(4\epsilon_{{\bf k}})$.


\begin{thebibliography}{99}
\bibitem{Boyd2008}R. W. , Nonlinear Optics, 3rd ed. (Academic Press,
Inc., Orlando, 2008).

\bibitem{Tokman2019}M. Tokman, S. B. Bodrov, Y. A. Sergeev, A. I.
Korytin, I. Oladyshkin, Y. Wang, A. Belyanin, and A. N. Stepanov,
Phys. Rev. B \textbf{99}, 155411 (2019).

\bibitem{Vandelli2019}M. Vandelli, M. I. Katsnelson, and E. A. Stepanov,
Phys. Rev. B \textbf{99}, 165432 (2019).

\bibitem{Ikeda2020}T. N. Ikeda, Phys. Rev. Research \textbf{2}, 032015
(2020).

\bibitem{Zhang2022}M. Zhang, N. Han, J. Wang, Z. Zhang, K. Liu, Z.
Sun, J. Zhao, and X. Gan, Nano Lett. \textbf{22}, 4287 (2022). 

\bibitem{Janisch2014}C. Janisch, Y. Wang, D. Ma, N. Mehta, A. L.
Elías, N. Perea-López, M. Terrones, V. Crespi, and Z. Liu, Sci Rep
\textbf{4}, 5530 (2014).

\bibitem{Rosa2018}H. G. Rosa, Y. Wei Ho, I. Verzhbitskiy, M. J. F.
L. Rodrigues, T. Taniguchi, K. Watanabe, G. Eda, V. M. Pereira, and
J. C. V. Gomes, Sci. Rep. \textbf{8}, 10035 (2018). 

\bibitem{Ho2020}Y. W. Ho, H. G. Rosa, I. Verzhbitskiy, M. J. L. F.
Rodrigues, T. Taniguchi, K. Watanabe, G. Eda, V. M. Pereira, and J.
C. Viana-Gomes, ACS Photonics \textbf{7}, 925 (2020).

\bibitem{Kuang2021}J. Yu, X. Kuang, J. Li, J. Zhong, C. Zeng, L.
Cao, Z. Liu, Z. Zeng, Z. Luo, T. He, A. Pan, and Y. Liu, Nat Commun
\textbf{12}, 1083 (2021). 

\bibitem{Mouchliadis2021}L. Mouchliadis, S. Psilodimitrakopoulos,
G. M. Maragkakis, I. Demeridou, G. Kourmoulakis, A. Lemonis, G. Kioseoglou,
and E. Stratakis, Npj 2D Mater Appl \textbf{5}, 6 (2021).

\bibitem{Wu2017}L. Wu, S. Patankar, T. Morimoto, N. L. Nair, E. Thewalt,
A. Little, J. G. Analytis, J. E. Moore, and J. Orenstein, Nature Phys
\textbf{13}, 350 (2017). 

\bibitem{Abulikemu2021}A. Abulikemu, Y. Kainuma, T. An, and M. Hase,
ACS Photonics 8, 988 (2021).

\bibitem{Baltz1981}R. von Baltz and W. Kraut, Phys. Rev. B \textbf{23},
5590 (1981). 

\bibitem{Cook2017}A. M. Cook, B. M. Fregoso, F. de Juan, S. Coh,
and J. E. Moore, Nat Commun \textbf{8}, 14176 (2017).

\bibitem{Tan2016}L. Z. Tan and A. M. Rappe, Phys. Rev. Lett. \textbf{116},
237402 (2016).

\bibitem{Ishizuka2022}H. Ishizuka and M. Sato, Phys. Rev. Lett. \textbf{129},
107201 (2022). 

\bibitem{Alon1998}{[}1{]} O. E. Alon, V. Averbukh, and N. Moiseyev,
Phys. Rev. Lett. \textbf{80}, 3743 (1998). 

\bibitem{Morimoto2017}T. Morimoto, H. C. Po, and A. Vishwanath, Phys.
Rev. B 95, 195155 (2017). 

\bibitem{Neufeld2019}O. Neufeld, D. Podolsky, and O. Cohen, Nat Commun
\textbf{10}, 405 (2019).

\bibitem{Ceccherini2001}F. Ceccherini, D. Bauer, and F. Cornolti,
J. Phys. B: At. Mol. Opt. Phys. \textbf{34}, 5017 (2001).

\bibitem{Fregoso2013}B. M. Fregoso, Y. H. Wang, N. Gedik, and V.
Galitski, Phys. Rev. B \textbf{88}, 155129 (2013).

\bibitem{Kuehn2009}W. Kuehn, K. Reimann, M. Woerner, and T. Elsaesser,
The Journal of Chemical Physics \textbf{130}, 164503 (2009). 

\bibitem{Kuehn2011}W. Kuehn, K. Reimann, M. Woerner, T. Elsaesser,
and R. Hey, J. Phys. Chem. B \textbf{115}, 5448 (2011).

\bibitem{Engel2007}G. S. Engel, T. R. Calhoun, E. L. Read, T.-K.
Ahn, T. Man\v{c}al, Y.-C. Cheng, R. E. Blankenship, and G. R. Fleming,
Nature 446, 782 (2007). 

\bibitem{Mahmood2021}F. Mahmood, D. Chaudhuri, S. Gopalakrishnan,
R. Nandkishore, and N. P. Armitage, Nat. Phys. \textbf{17}, 627 (2021). 

\bibitem{Lu2017}J. Lu, X. Li, H. Y. Hwang, B. K. Ofori-Okai, T. Kurihara,
T. Suemoto, and K. A. Nelson, Phys. Rev. Lett. \textbf{118}, 207204
(2017).

\bibitem{Wan2019}Y. Wan and N. P. Armitage, Phys. Rev. Lett. \textbf{122},
257401 (2019).

\bibitem{Li2021}Z.-L. Li, M. Oshikawa, and Y. Wan, Phys. Rev. X \textbf{11},
031035 (2021). 

\bibitem{Nandkishore2021}R. M. Nandkishore, W. Choi, and Y. B. Kim,
Phys. Rev. Research \textbf{3}, 013254 (2021).

\bibitem{Choi2020}W. Choi, K. H. Lee, and Y. B. Kim, Phys. Rev. Lett.
\textbf{124}, 117205 (2020).

\bibitem{Qiang2023}Y. Qiang, V. L. Quito, T. V. Trevisan, and P.
P. Orth, arXiv:2301.11243. 

\bibitem{Savary2016}L. Savary and L. Balents, Rep. Prog. Phys. \textbf{80},
016502 (2016).

\bibitem{Kitaev2006}A. Kitaev, Annals of Physics \textbf{321}, 2
(2006).

\bibitem{Takagi2019}H. Takagi, T. Takayama, G. Jackeli, G. Khaliullin,
and S. E. Nagler, Nat Rev Phys \textbf{1}, 264 (2019).

\bibitem{Motome2019}Y. Motome and J. Nasu, J. Phys. Soc. Jpn. \textbf{89},
012002 (2019).

\bibitem{Jackeli2009}G. Jackeli and G. Khaliullin, Phys. Rev. Lett.
\textbf{102}, 017205 (2009).

\bibitem{Baskaran2007}G. Baskaran, S. Mandal, and R. Shankar, Phys.
Rev. Lett. \textbf{98}, 247201 (2007).

\bibitem{Plumb2014}K. W. Plumb, J. P. Clancy, L. J. Sandilands, V.
V. Shankar, Y. F. Hu, K. S. Burch, H.-Y. Kee, and Y.-J. Kim, Phys.
Rev. B \textbf{90}, 041112 (2014).

\bibitem{Cao2016}H. B. Cao, A. Banerjee, J.-Q. Yan, C. A. Bridges,
M. D. Lumsden, D. G. Mandrus, D. A. Tennant, B. C. Chakoumakos, and
S. E. Nagler, Phys. Rev. B \textbf{93}, 134423 (2016). 

\bibitem{Sears2017}J. A. Sears, Y. Zhao, Z. Xu, J. W. Lynn, and Y.-J.
Kim, Phys. Rev. B \textbf{95}, 180411 (2017). 

\bibitem{Anja2017}A. U. B. Wolter, L. T. Corredor, L. Janssen, K.
Nenkov, S. Sch\"onecker, S.-H. Do, K.-Y. Choi, R. Albrecht, J. Hunger,
T. Doert, M. Vojta, and B. B\"uchner, Phys. Rev. B \textbf{96}, 041405
(2017).

\bibitem{Baek2017}S.-H. Baek, S.-H. Do, K.-Y. Choi, Y. S. Kwon, A.
U. B. Wolter, S. Nishimoto, J. van den Brink, and B. B\"uchner, Phys.
Rev. Lett. \textbf{119}, 037201 (2017)

\bibitem{Hentrich2018}R. Hentrich, A. U. B. Wolter, X. Zotos, W.
Brenig, D. Nowak, A. Isaeva, T. Doert, A. Banerjee, P. Lampen-Kelley,
D. G. Mandrus, S. E. Nagler, J. Sears, Y.-J. Kim, B. B\"uchner, and
C. Hess, Phys. Rev. Lett. \textbf{120}, 117204 (2018).

\bibitem{Balz2019}C. Balz, P. Lampen-Kelley, A. Banerjee, J. Yan,
Z. Lu, X. Hu, S. M. Yadav, Y. Takano, Y. Liu, D. A. Tennant, M. D.
Lumsden, D. Mandrus, and S. E. Nagler, Phys. Rev. B \textbf{100},
060405 (2019).

\bibitem{Schonemann2020}R. Sch\"onemann, S. Imajo, F. Weickert,
J. Yan, D. G. Mandrus, Y. Takano, E. L. Brosha, P. F. S. Rosa, S.
E. Nagler, K. Kindo, and M. Jaime, Phys. Rev. B \textbf{102}, 214432
(2020).

\bibitem{Balz2021}C. Balz, L. Janssen, P. Lampen-Kelley, A. Banerjee,
Y. H. Liu, J.-Q. Yan, D. G. Mandrus, M. Vojta, and S. E. Nagler, Phys.
Rev. B \textbf{103}, 174417 (2021).

\bibitem{Knolle2014}J. Knolle, D. L. Kovrizhin, J. T. Chalker, and
R. Moessner, Phys. Rev. Lett. \textbf{112}, 207203 (2014). 

\bibitem{Banerjee2016}{[}1{]} A. Banerjee, C. A. Bridges, J.-Q. Yan,
A. A. Aczel, L. Li, M. B. Stone, G. E. Granroth, M. D. Lumsden, Y.
Yiu, J. Knolle, S. Bhattacharjee, D. L. Kovrizhin, R. Moessner, D.
A. Tennant, D. G. Mandrus, and S. E. Nagler, Nature Materials \textbf{15},
733 (2016). 

\bibitem{Banerjee2017}A. Banerjee, J. Yan, J. Knolle, C. A. Bridges,
M. B. Stone, M. D. Lumsden, D. G. Mandrus, D. A. Tennant, R. Moessner,
and S. E. Nagler, Science \textbf{356}, 1055 (2017).

\bibitem{Do2017}S.-H. Do, S.-Y. Park, J. Yoshitake, J. Nasu, Y. Motome,
Y. S. Kwon, D. T. Adroja, D. J. Voneshen, K. Kim, T.-H. Jang, J.-H.
Park, K.-Y. Choi, and S. Ji, Nature Phys \textbf{13}, 1079 (2017). 

\bibitem{Sandilands2015}L. J. Sandilands, Y. Tian, K. W. Plumb, Y.-J.
Kim, and K. S. Burch, Phys. Rev. Lett. \textbf{114}, 147201 (2015). 

\bibitem{Nasu2016}J. Nasu, J. Knolle, D. L. Kovrizhin, Y. Motome,
and R. Moessner, Nature Physics \textbf{12}, 912 (2016). 

\bibitem{Wulferding2020}D. Wulferding, Y. Choi, S.-H. Do, C. H. Lee,
P. Lemmens, C. Faugeras, Y. Gallais, and K.-Y. Choi, Nat Commun \textbf{11},
1 (2020). 

\bibitem{Halasz2016}G. B. Hal\'asz, N. B. Perkins, and J. van den
Brink, Phys. Rev. Lett. \textbf{117}, 127203 (2016).

\bibitem{Metavitsiadis2020}A. Metavitsiadis and W. Brenig, Phys.
Rev. B \textbf{101}, 035103 (2020).

\bibitem{Metavitsiadis2022}A. Metavitsiadis, W. Natori, J. Knolle,
and W. Brenig, Phys. Rev. B \textbf{105}, 165151 (2022).

\bibitem{Ye2020}M. Ye, R. M. Fernandes, and N. B. Perkins, Phys.
Rev. Research \textbf{2}, 033180 (2020). 

\bibitem{Feng2021}K. Feng, M. Ye, and N. B. Perkins, Phys. Rev. B
\textbf{103}, 214416 (2021).

\bibitem{Feng2022}K. Feng, S. Swarup, and N. B. Perkins, Phys. Rev.
B \textbf{105}, L121108 (2022).

\bibitem{Li_diff_2021}H. Li, T. T. Zhang, A. Said, G. Fabbris, D.
G. Mazzone, J. Q. Yan, D. Mandrus, G. B. Halász, S. Okamoto, S. Murakami,
M. P. M. Dean, H. N. Lee, and H. Miao, Nat Commun \textbf{12}, 3513
(2021). 

\bibitem{Hauspurg2023}A. Hauspurg, S. Zherlitsyn, T. Helm, V. Felea,
J. Wosnitza, V. Tsurkan, K.-Y. Choi, S.-H. Do, M. Ye, W. Brenig, and
N. B. Perkins, arXiv:2303.09288. 

\bibitem{Kanega2021}M. Kanega, T. N. Ikeda, and M. Sato, Phys. Rev.
Research \textbf{3}, L032024 (2021). 

\bibitem{Katsura2009}H. Katsura, M. Sato, T. Furuta, and N. Nagaosa,
Phys. Rev. Lett. \textbf{103}, 177402 (2009).

\bibitem{Lorenzana1995}J. Lorenzana and G. A. Sawatzky, Phys. Rev.
Lett. \textbf{74}, 1867 (1995).

\bibitem{Jurecka2000}C. Jurecka and W. Brenig Phys. Rev. B \textbf{61},
14307 (2000).

\bibitem{Tokura2014}Y. Tokura, S. Seki, and N. Nagaosa, Rep. Prog.
Phys. \textbf{77}, 076501 (2014).

\bibitem{Symmetry}Generalizing field directions is postponed to future
work.

\bibitem{Aktsipetrov1996}O. A. Aktsipetrov, A. A. Fedyanin, E. D.
Mishina, A. N. Rubtsov, C. W. van Hasselt, M. A. C. Devillers, and
Th. Rasing, Phys. Rev. B 54, 1825 (1996).

\bibitem{Bykov2012}A. Y. Bykov, T. V. Murzina, M. G. Rybin, and E.
D. Obraztsova, Phys. Rev. B 85, 121413 (2012).

\bibitem{Nasu2015}J. Nasu, M. Udagawa, and Y. Motome, Phys. Rev.
B \textbf{92}, 115122 (2015).

\bibitem{Metavitsiadis2017}A. Metavitsiadis, A. Pidatella, and W.
Brenig, Phys. Rev. B \textbf{96}, 205121 (2017).

\bibitem{Pidatella2019}A. Pidatella, A. Metavitsiadis, and W. Brenig
Phys. Rev. B \textbf{99}, 075141 (2019).

\bibitem{RndSec}The term ``random flux sector'' is used to imply
an average over sectors with fixed, random values of $\text{\ensuremath{\eta_{\mathbf{l}}}}$.

\bibitem{Metavitsiadis2017a}A. Metavitsiadis and W. Brenig, Rev.
B \textbf{96}, 041115(R) (2017).

\bibitem{Metavitsiadis2021}A. Metavitsiadis and W. Brenig, Phys.
Rev. B 104, 104424 (2021). 

\bibitem{leadord}This is merely for simplicity. No conceptual or
technical reason prevents inclusion of graphs beyond the leading order
in $E_{ac}(t)$, at the expense of increasing the length of the analytical
expressions.

\bibitem{higo}Upon reasonable request, $\chi_{N\omega}(\omega)$
for $N>3$ will be provided in private communication.

\bibitem{C3symm}The relative orientation of the polarization in Eq.
(\ref{eq:P}) and the gauge-fixing on the $z$-bonds does not imply
unphysical anisotropies. This is satisfied by our $\mathbf{r}$-space
code. I.e. the spectra we obtain in the random gauge sector are independent
of $(P,E)$ pointing perpendicular to the $z$, $x$, or $y$ bonds.

\bibitem{Fetter1971}A. L. Fetter and J. D. Walecka, \emph{Quantum
Theory of Many-Particle Systems}, McGraw-Hill, Boston, (1971)

\bibitem{Butcher1990}P. N. Butcher, D. Cotter, The Elements of Nonlinear
Optics, Cambridge University Press, Cambridge, 1990.

\bibitem{Evans}W. A. B. Evans, Proc. Phys. Soc. \textbf{88}, 723
(1966).

\bibitem{Rostami}H. Rostami, M. I. Katsnelson, G. Vignale, and M.
Polini, Annals of Physics \textbf{431}, 168523 (2021).

\bibitem{wbunp}W. Brenig, unpublished.

\end{thebibliography}
\end{document}